\documentclass{jnmp}

\usepackage{amsmath}
\usepackage{graphicx}

\JNMPnumberwithin{equation}{section}

\begin{document}

\renewcommand{\evenhead}{F~Calogero and M~Sommacal}
\renewcommand{\oddhead}{Periodic Solutions of a System of Complex ODEs. II. Higher Periods}

\thispagestyle{empty}

\FirstPageHead{9}{4}{2002}{\pageref{calogero-firstpage}--\pageref{calogero-lastpage}}{Article}

\copyrightnote{2002}{F~Calogero and M~Sommacal}

\Name{Periodic Solutions of a System of Complex ODEs.\\ II. Higher Periods}
\label{calogero-firstpage}

\Author{F~CALOGERO~$^{\dag\,\ddag}$ and M~SOMMACAL~$^\ddag$}

\Address{$^\dag$~Dipartimento di Fisica, Universit\'a di Roma ``La Sapienza'', 00185 Roma, Italy\\[10pt]
$^\ddag$ Istituto Nazionale di Fisica Nucleare, Sezione di Roma, Italy\\[5pt]
~~E-mail: francesco.calogero@roma1.infn.it, \ \ francesco.calogero@uniroma1.it \\
~~E-mail: matteo.sommacal@roma1.infn.it}

\Date{Received March 4, 2002; Accepted April 5, 2002}

\begin{abstract}
\noindent
In a previous paper the \textit{real} evolution of the system of ODEs
\begin{gather*}
\ddot{z}_{n} + z_{n}=\sum\limits_{m = 1,\; m \ne n}^{N} g_{nm}{(z_{n} - z_{m})}
^{- 3},\\
 z_{n} \equiv z_{n}(t), \qquad \dot {z}_{n} \equiv \frac{d z_{n}(t)}{dt},
\qquad  n = 1,\ldots,N
\end{gather*}
is discussed in $C_{N} $, namely the $N$ dependent variables $z_{n} $, as
well as the $N( {N - 1} )$ (arbitrary!) ``coupling constants''
$g_{nm} $, are considered to be \textit{complex} numbers, while the
independent variable $t$ (``time'') is \textit{real}. In that context it was
proven that there exists, in the phase space of the initial data $z_{n}
(0)$, $\dot {z}_{n} (0)$, an open domain having
\textit{infinite} measure, such that \textit{all} trajectories emerging from
it are \textit{completely periodic} with period $2\pi $, $z_{n} (t + 2\pi)
= z_{n} (t)$. In this paper we investigate,
both by analytical techniques and via the display of numerical simulations,
the remaining solutions, and in particular we show that there exist many ---
emerging out of sets of initial data having nonvanishing measures in the
phase space of such data --- that are also \textit{completely periodic} but
with periods which are \textit{integer multiples} of $2\pi $. We also
elucidate the mechanism that yields \textit{nonperiodic} solutions,
including those characterized by a ``chaotic'' behavior, namely those
associated, in the context of the initial-value problem, with a
\textit{sensitive dependence} on the initial data.
\end{abstract}

\section{Introduction}

In a previous paper [1] one of us (FC) analyzed, in the \textit{complex}
domain, the dynamical system characterized by the Newtonian equations of
motion
\begin{gather}
\ddot{z}_{n} + z_{n}=\sum_{m = 1,\; m \ne n}^{N} g_{nm}{(z_{n} -
z_{m})}
^{- 3} , \nonumber\\
z_{n} \equiv z_{n}(t) ,\qquad  \dot {z}_{n} \equiv \frac{d z_{n}(t)}{dt},\qquad n
= 1,\ldots,N\label{eq1}
\end{gather}
which obtain in the standard manner from the Hamiltonian
\begin{subequations}
\begin{equation}\label{eq2}
H( \underline {z} ,\underline {p}) = \frac{1}{2}\sum\limits_{n =
1}^{N} \left( p_{n}^{2} + z_{n}^{2}\right) +
\frac{1}{4}\sum\limits_{n,m = 1,\; n \ne m}^{N} {} g_{nm} (z_{n} -
z_{m})^{- 2},
\end{equation}
provided (as we hereafter assume --- even though the main result of [1] holds
without this restriction)
\begin{equation}\label{eq3}
g_{nm} = g_{mn}.
\end{equation}
\end{subequations}
Here and below $N$ is an arbitrary positive integer ($N \ge 2$), the indices
$n$, $m$ run from $1$ to $N$ unless otherwise indicated, underlined
quantities are $N$-vectors, say $\underline {z} \equiv (z_{1} ,\ldots,z_{N})$,
and all quantities (namely, the $N$ ``canonical
coordinates'' $z_{n} $, the $N$ ``canonical momenta'' $p_{n} $, the $N( N - 1)/2$
``coupling constants'' $g_{nm} $) are \textit{complex},
while the independent variable~$t$ (``time'') is instead \textit{real}. Of
course the $N$ complex equations of motion (\ref{eq1}) can be reformulated [1] as
$2N$ real~--- and as well Hamiltonian [2]~--- equations of motion, by introducing
the real and imaginary parts of the coordinates $z_{n} $, $z_{n} \equiv
x_{n} + i y_{n} $, or their amplitudes and phases, $z_{n} \equiv \rho _{n}
\exp(i\theta_{n})$:
\begin{subequations}
\begin{gather}
\ddot{x}_{n}+x_{n} = \sum \limits_{m=1 , \; m \ne n}^{N}
r_{nm}^{-6}
\left[a_{nm}x_{nm}\left(x_{nm}^{2}-3y_{nm}^{2}\right)-b_{nm}y_{nm}
\left(y_{nm}^{2}-3x_{nm}^{2}\right) \right] \nonumber\\
\phantom{\ddot{x}_{n}+x_{n}}{}= \sum \limits_{m=1 , \; m \ne
n}^{N} r_{nm}^{-3}
|g_{nm}| \cos(\gamma_{nm}-3\theta_{nm}), \label{eq3bis}\\
\ddot{y}_{n}+y_{n} = \sum \limits_{m=1, \; m \ne n}^{N}
r_{nm}^{-6} \left[a_{nm}y_{nm}\left(y_{nm}^{2}-3x_{nm}^{2}\right)
-b_{nm}y_{nm}\left(x_{nm}^{2}-3y_{nm}^{2}\right) \right] \nonumber\\
\phantom{\ddot{y}_{n}+y_{n}}{}= \sum \limits_{m=1 , \; m \ne
n}^{N} r_{nm}^{-3} |g_{nm}| \sin(\gamma_{nm}-3\theta_{nm}),
\label{eq3ter}\end{gather} where of course
\begin{gather}
z_{n} = x_{n}+i y_{n}, \nonumber\\
z_{nm} \equiv z_{n}-z_{m} = x_{n}-x_{m}+i (y_{n}-y_{m}) \equiv
x_{nm} + i y_{nm} = r_{nm} \exp (i \theta_{nm}) , \nonumber\\
g_{nm} = a_{nm} + i b_{nm} = |g_{nm}| \exp (i \gamma_{nm}).\label{eq3quater}
\end{gather}
\end{subequations}
We shall return to the motivations for this choice to investigate the system
(\ref{eq1}) in the complex domain at the end of this Section~1.

If all the coupling constants coincide, $g_{nm} = g$, the
Hamiltonian system (\ref{eq1}) is a~well-known \textit{completely
integrable} many-body model (see for instance~[2] and the
references quoted there), and \textit{all} its nonsingular
solutions are \textit{completely periodic} with period $2\pi $, or
possibly an integer multiple of $2\pi $. (Indeed, in this
\textit{integrable} case the $N$ coordinates~$z_{n}(t)$ can be
identified with the $N$ zeros of a polynomial of degree $N$ the
coefficients of which are periodic in $t$ with period $2\pi $, so
that the set of these $N$ zeros is also periodic with period $2\pi
$, and each individual zero is therefore also periodic, although
possibly with a~larger period which is an integer multiple of
$2\pi $ due to a possible reshuffling of the zeros as the motion
unfolds; in the \textit{real} case with all coupling constants
\textit{equal} and \textit{positive}, $g_{nm} = g > 0$, when the
motions are confined to the real axis and no such reshuffling can
occur due to the singular and repulsive character of the two-body
forces, \textit{all} real solutions are \textit{nonsingular} and
\textit{completely periodic} with period $2\pi$, $\underline {z}
(t + 2\pi) = \underline {z} (t)$; see for instance~[2]). Here we
focus instead on the more general case with \textit{completely
arbitrary} coupling constants $g_{nm} $, which is generally
believed \textit{not} to be integrable. But even in this case ---
as proven in~[1]~--- there does exist a domain of initial data
$\underline {z} (0)$, $\underline {\dot {z}} (0)$ having
\textit{infinite} measure in phase space~--- indeed, having a
measure which is a finite fraction of that of the entire phase
space --- such that \textit{all} the trajectories originating from
it are \textit{completely periodic} with period $2\pi $,
$\underline {z} (t + 2\pi) = \underline {z} (t)$. As pointed out
in~[1], this is a somewhat surprising finding, inasmuch as it
negates the expectation that, for a nonlinear dynamical system
with several degrees of freedom that possesses completely periodic
trajectories emerging from some specific initial data, any generic
variation of these initial data destroy the complete periodicity
of the trajectories or at least change their period
--- unless the system is essentially equivalent (say, via an appropriate
change of variables) to a linear system (such as (\ref{eq1}) with \textit{all}
coupling constants vanishing, $g_{nm} = 0$), which is certainly not the case
for the system (\ref{eq1}) with \textit{arbitrary} coupling constants $g_{nm}$,
at least not in any manner explicitly computable in closed form.

But, as shown in [1], this fact is a rather elementary consequence of an
approach (a ``trick'') introduced and rather extensively used recently to
evince analogous results (see~[2] and the references quoted there, as well
as [3, 4, 5]). This same trick can as well be exploited to investigate
the remaining solutions, namely those not belonging to the class of
\textit{completely periodic} solutions with period $2\pi $ the existence
of which was proven in~[1]. This we do in the present paper, and we also
confirm the insight thereby gained by exhibiting numerical solutions of
(\ref{eq1}) performed via a computer code created by one of us (MS)~[6]. In
particular we demonstrate below the existence of open domains of initial
data, having \textit{nonvanishing} measures in the phase space of such data,
which also yield \textit{completely periodic} solutions, but with periods
which are \textit{integer} \textit{multiples} of $2\pi $, and we also
elucidate the mechanism that originates \textit{non-periodic} and \textit{chaotic}
solutions.

In the next Section (the first part of which is closely patterned after
Section~2 of [1] and is reported here to make this paper self-contained) the
``trick'' mentioned above~--- which in fact amounts to a change of (dependent
and independent) variables --- is introduced. In Section~3 the analyticity
properties in a time-like complex variable are investigated of the solutions
of the system of ODEs, see (\ref{eq11}), related to (\ref{eq1}) via the trick. The
implications of these findings as regards the periodicity of certain
solutions of (\ref{eq1}) are discussed in Section~4; in particular the mechanism
that underlies the existence of \textit{completely periodic} motions
with \textit{higher} periods (\textit{integer multiples} of $2\pi $), and of
\textit{non-periodic} and \textit{chaotic} motions, is elucidated. In Section~5 numerical
examples of these trajectories are exhibited. In Section~6 some final
remarks are proffered. Two appendices complete our presentation: Appendix~A
contains some developments (confined there not to interrupt
the flow of presentation in Section~3) concerning the analytic structure of the
solutions of the evolution equations~(\ref{eq11}); Appendix~B focuses on the
\textit{three-body} case ($N = 3$), in particular it reports its
reducibility to quadratures and a discussion of the information about the
analytic behavior of the solutions of (\ref{eq11}) entailed by this fact.

Let us end this introductory Section 1 with some remarks on the
choice to investigate the motion determined by the Newtonian
equations~(\ref{eq1}) in the \textit{complex}, rather than the
\textit{real}, domain. A clear hint that, at least from
a~mathematical point of view, this is a~more natural environment
to work in, comes already from the treatment of (\ref{eq1}) in the
\textit{integrable} case with \textit{equal} coupling constants,
$g_{nm}=g$, since, as mentioned above, it is then appropriate to
identify the $N$ particle positions $z_{n} \equiv z_{n}(t)$ with
the $N$ zeros of a time dependent (monic) polynomial of degree $N$
in $z$, say $P_{N}(z,t)$ such that $P_{N} [ t, z_{n}(t)] = 0$ (see
for instance Ref.~[2]); and clearly the natural environment to
investigate the zeros of a polynomial is the \textit{complex}
plane rather than the \textit{real} line. In our context, an
essential motivation to work in the complex comes from the
important role that analyticity properties play in our treatment,
see below. Moreover motions roaming over the complex plane display
a much richer dynamics than those restricted to the real line,
especially in the case with singular interparticle forces, because
of the possibility in the former case, but not in the latter, that
particles go around each other. And it is then natural to
re-interpret the (\textit{complex}) $N$-body problem (\ref{eq1})
as describing the (\textit{real}) motion of $N$ particles
(\textit{in the plane}), by introducing a one-to-one
correspondence among the complex coordinates $z_{n} \equiv x_{n} +
i y_{n}$, see~(\ref{eq3quater}), and the real two-vectors in the
plane $\vec{r}_{n} \equiv (x_{n}, y_{n})$. But this approach, that
is quite convenient to identify interesting many-body problems in
the plane (see Chapter~4 of Ref.~[2]), suffers in the present case
from a drawback: the resulting many-body problem in the plane is
not rotation-invariant, see~(1.3). Indeed the many-body
problem~(1.3) is characterized by a (clearly rotation-invariant)
harmonic-oscillator one-body force attracting every particle
towards the origin, and by a (clearly not rotation-invariant)
singular two-body force acting among each particle pair. As
clearly seen from~(1.3), the strength of the two-body force is
proportional to the inverse cube of the (Euclidian) interparticle
distance, hence it generally diverges when two particles collide;
but this force also depends, both in modulus and direction, from
the orientation of the interparticle vector, as well as from the
phase $\gamma_{nm}$ of the relevant interparticle coupling
constant (\ref{eq3quater}). Indeed the two-body force
$\vec{f}_{nm}$ acting on the $n$-th particle due to the $m$-th
particle is the two-vector $\vec{f}_{nm}=r_{nm}^{-3}|g_{nm}| (
\cos(\gamma_{nm}-3\theta_{nm}), \sin(\gamma_{nm}-3\theta_{nm}) )$,
which is generally not aligned to the interparticle distance
$\vec{r}_{nm}= r_{nm}(\cos\theta_{nm}, \sin\theta_{nm})$,
see~(1.3). The diligent reader is advised to try and become
familiar with the specific implications of this fact, as they will
be eventually helpful to understand the trajectories of the
solutions of the model (\ref{eq1}), see Section~5.

\section{The trick}

In this section we describe the ``trick'' [1--4] that underlies our subsequent
findings, and at the end we mention a remarkable property of the system of
ODEs (\ref{eq1}). But firstly we rewrite, mainly for notational convenience, these
equations of motion, (\ref{eq1}), as follows:
\begin{subequations}
\begin{equation}
\label{eq4}
\ddot {z}_{n} + \omega ^{2} z_{n} = \sum\limits_{m = 1,\; m \ne n}^{N}
g_{nm} ( z_{n} - z_{m})^{- 3},
\end{equation}
and we note that the corresponding Hamiltonian reads
\begin{equation}
\label{eq5}
H(\underline {z} ,\underline {p}) =
\frac{1}{2}\sum\limits_{n = 1}^{N} \left( p_{n}^{2} + \omega
^{2}z_{n}^{2}\right) + \frac{1}{4}\sum\limits_{n,m = 1,\; n \ne
m}^{N} g_{nm} (z_{n} - z_{m})^{- 2}.
\end{equation}
\end{subequations}
Here we introduce the additional constant $\omega $, which is hereafter
assumed to be \textit{positive}, $\omega > 0$, and to which we associate the
basic period
\begin{equation}
\label{eq6}
T = 2\pi /\omega .
\end{equation}
In the following it will sometimes be convenient to set $\omega = 1$ so that
(\ref{eq4}) coincide with (\ref{eq1}) and the basic period becomes $T = 2\pi $, or to
set instead $\omega = 2\pi $ so that the basic period becomes unity, $T =
1$. Of course these cases are all related via a rescaling of the dependent
and independent variables, since clearly by setting
\begin{subequations}
\begin{equation}\label{eq7}
\tilde {z}(\tilde t) = az(t),\qquad
\tilde {t} = bt,\qquad
\tilde {\omega}  = \omega /b,\qquad
\tilde {g}_{nm} = \left(a^{2}/b \right)^{2}g_{nm}
\end{equation}
with $a$, $b$ two \textit{positive} rescaling constants which can be chosen
at our convenience, the ODEs (\ref{eq4}) get reformulated in a completely
analogous ``tilded'' version,
\begin{equation}\label{eq8}
\tilde {z}''_{n} + \tilde {\omega} ^{2} \tilde {z}_{n} = \sum\limits_{m =
1,\; m \ne n}^{N} \tilde {g}_{nm} ( \tilde {z}_{n} - \tilde {z}_{m}
)^{- 3},
\end{equation}
\end{subequations}
where of course here the primes indicate differentiations with respect to
the argument of the function they are appended to, $\tilde{z}' \equiv
d\tilde {z}(\tilde t)/d\tilde {t}$. Note in
particular that by setting $b = a^{2} = \omega $ one gets $\tilde {\omega
} = 1$, $\tilde {g}_{nm} = g_{nm} $, namely the tilded version (\ref{eq8})
reproduces essentially (\ref{eq1}).

Now, the ``trick''. Let us set
\begin{subequations}
\begin{gather}
\label{eq9}
z_{n} (t) = \exp(  - i\omega t)\zeta _{n}(\tau),\\
\label{eq10}
\tau \equiv \tau (t) = [\exp(2i\omega t) - 1]/(2i\omega).
\end{gather}
\end{subequations}
As can be readily verified, this change of (dependent and independent)
variables, (2.4), transforms (\ref{eq4}) into
\begin{equation}
\label{eq11}
\zeta''_{n} = \sum\limits_{m = 1,\; m \ne n}^{N} g_{nm} (\zeta
_{n} - \zeta _{m})^{- 3}.
\end{equation}
Here and below appended primes denote derivatives with respect to the new
independent variable $\tau $, while of course the dots in (\ref{eq4}) and below
denote as usual derivatives with respect to the \textit{real} time~$t$.

The change of variables (2.4) entails the following relations among the
initial data, $\underline {z} (0)$,
$\underline {\dot {z}} (0)$, respectively $\underline {\zeta}  (0)$,
$\underline {\zeta '} (0)$, for (\ref{eq4})
respectively (\ref{eq11}):
\begin{subequations}
\begin{gather}
\label{eq12}
z_{n} (0) = \zeta _{n} (0),\\
\label{eq13}
\dot {z}_{n} (0) = \zeta'_{n} (0) - i\omega
\zeta _{n} (0).
\end{gather}
\end{subequations}

We now observe that, as the (real, ``physical time'') variable $t$
varies from $0$ to $T/2 = \pi /\omega $, the (complex) variable
$\tau $ travels (counterclockwise) full circle over the circular
contour~$\tilde {C}$, the diameter of which, of length $1/\omega =
T/(2\pi)$, lies on the upper-half of the complex $\tau $-plane,
with its lower end at the origin, $\tau = 0$, and its upper end at
$\tau = i/\omega $. Hence if the solution $\underline {\zeta}
(\tau)$ of (\ref{eq11}) which emerges from some assigned initial
data $\underline {\zeta}  (0)$, $\underline {\zeta '} (0)$ is
\textit{holomorphic}, as a ($N$-vector-valued) function of the
complex variable $\tau$, in the closed circular disk $C$ encircled
by the circle $\tilde {C}$ in the complex $\tau$-plane, then the
corresponding solution $\underline {z} (t)$ of (\ref{eq4}),
related to $\underline {\zeta}  (\tau)$ by (2.4), is
\textit{completely periodic} in $t$ with period $T$, $\underline
{z} (t + T) = \underline {z} (t)$ (see (\ref{eq6}); and note that
$\underline {\zeta}  (\tau)$, considered as function of the
\textit{real} va\-ria\-ble $t$, is then \textit{periodic} with
period $T/2$, but $\underline {z}(t)$ is instead
\textit{antiperiodic} with period~$T/2$, $\underline {z} (t + T/2)
= - \underline {z} (t)$, due to the prefactor $\exp( - i\omega
t)$, see (\ref{eq9})).

In [1] it was proven that there indeed exists a domain, having
\textit{infinite} measure in phase space, of initial data
$\underline {\zeta } (0)$, $\underline {\zeta'} (0)$ such that the
corresponding solutions $\underline {\zeta}  (\tau)$ of
(\ref{eq11}) are \textit{holomorphic} in $\tau $ in the disk
$C$~--- and this fact implies the existence of an open domain,
having as well \textit{infinite} measure in phase space, of
corresponding initial data $\underline {z} (0)$, $\underline {\dot
{z}} (0)$ such that the corresponding solutions $\underline {z}
(t)$ of (\ref{eq4}) are \textit{completely periodic} with period
$T$, $\underline {z} (t + T) = \underline {z} (t)$. In the
following Section~3 we show that the singularities of the
solutions~$\underline {\zeta}  (\tau)$ of (\ref{eq11})~---
considered as functions of the complex variable $\tau $ --- are
\textit{branch points} of \textit{square-root} type, and in
Section~4 we infer from this that, whenever the
solution~$\underline {\zeta}  (\tau)$ of~(\ref{eq11}) has a
\textit{finite} number of such branch points \textit{inside} the
circle~$\tilde {C}$~--- generally nested inside each other, namely
occurring on different Riemann sheets~--- then the corresponding
solution $\underline {z} (t)$ of (\ref{eq4}), considered as a
function of the \textit{real} ``time'' variable $t$, is again
\textit{completely periodic}, albeit now with a period which is an
\textit{integer multiple} of $T$. We also infer that when instead
the solution $\underline {\zeta}  (\tau)$ of (\ref{eq11}) has an
\textit{infinite} number of such \textit{square-root} branch
points \textit{inside} the circle $\tilde {C}$
--- again, generally nested inside each other, namely occurring on different
Riemann sheets~--- then the corresponding solution $\underline {z}
(t)$ of (\ref{eq4}), considered again as a function of the
\textit{real} ``time'' variable $t$, may be \textit{not periodic}
at all indeed it generally behaves \textit{chaotically} (actually,
strictly speaking, the \textit{chaotic character} is not a
property of a single solution, it rather has to do with the
\textit{difference} among the long-time\textit{} behavior of a
solution and those of \textit{other} solutions which emerge from
\textit{almost identical} initial data --- see below).

Finally let us report a remarkable property of the system of ODEs (\ref{eq4}).
First of all we recall a trivial result, namely that the center of mass of
the system (\ref{eq4}),
\begin{subequations}
\begin{equation}
\label{eq14}
\bar {z}(t) = N^{- 1}\sum\limits_{n = 1}^{N} z_{n} (t),
\end{equation}
rotates uniformly with period $T$, see (\ref{eq6}), since clearly these equations
of motion, (\ref{eq4}), entail
\begin{equation}
\label{eq15}
\ddot {\bar {z}} + \omega ^{2}\bar {z} = 0,
\end{equation}
hence
\begin{equation}
\label{eq16}
\bar {z}(t) = \bar {z}(0)\cos(\omega t) + \dot {\bar {z}}(0)
(\omega)^{- 1}\sin(\omega t).
\end{equation}
\end{subequations}

A less trivial result is that the sum of the squares of the particle coordinates,
\begin{subequations}
\begin{equation}
\label{eq17}
\bar {z}^{(2)}(t) = N^{- 1}\sum\limits_{n =
1}^{N} [ z_{n} (t)]^{2},
\end{equation}
evolves as well periodically, with period $T/2$. Indeed, as shown below, the
equations of motion (\ref{eq4}) entail
\begin{equation}
\label{eq18}
\ddot {\bar {z}}^{(2)} + (2\omega)^{2}\bar
{z}^{(2)} = (4/N)H,
\end{equation}
where the Hamiltonian $H$, see (\ref{eq5}), is of course a constant of motion.
Hence
\begin{equation}
\label{eq19}
\bar {z}^{(2)}(t) = a\exp(2i\omega t)+b+c\exp(-2 i\omega t),
\end{equation}
\end{subequations}
where the three constants $a$, $b$, $c$ are of course related to the initial
data as follows:
\begin{subequations}
\begin{gather}
\label{eq20}
\bar {z}(0) = a + b + c = N^{ - 1}\sum\limits_{n = 1}^{N}
[z_{n} (0)]^{2},\\
\label{eq21}
\dot {\bar {z}}(0) = 2i\omega (a - b) =
(2/N)\sum\limits_{n = 1}^{N} \dot {z}_{n} (0) z_{n} (0),\\
\label{eq22}
\ddot {\bar {z}}^{(2)} (0) = - (2\omega)^{2}(a + b) = - (2\omega)^{2}
\bar {z}^{(2)}(0) + (4/N)H.
\end{gather}
\end{subequations}

There remains to prove that the equations of motion (\ref{eq4}) entail (\ref{eq18}).
Indeed by differentiating twice the definition (\ref{eq17}) one
gets
\begin{subequations}
\begin{equation}
\label{eq23}
\ddot {\bar {z}}^{(2)} = \frac{2}{N}\sum\limits_{n =
1}^{N} \left( \dot {z}_{n}^{2} + z_{n} \ddot {z}_{n} \right),
\end{equation}
and by using the equations of motion (\ref{eq4}) this yields
\begin{equation}
\label{eq24}
\ddot {\bar {z}}^{(2)} = \frac{2}{N}\left[
\sum\limits_{n = 1}^{N}  \left( \dot {z}_{n}^{2} - \omega
^{2}z_{n}^{2}  \right) + \sum\limits_{n,m = 1;\; m \ne n}^{N} g_{nm}
z_{n} (z_{n} - z_{m})^{- 3}\right].
\end{equation}
Hence (using again the definition (\ref{eq17}), and the symmetry of the coupling
constants, see (\ref{eq3}))
\begin{equation}
\label{eq25}
\ddot {\bar {z}}^{(2)} + 4\omega ^{2}\bar {z}^{(2)}
= \frac{2}{N}\left[\sum\limits_{n = 1}^{N}
\left(\dot {z}_{n}^{2} + \omega ^{2}z_{n}^{2}  \right) + \frac 12
\sum\limits_{n,m = 1;\; m \ne n}^{N} g_{nm} (z_{n} - z_{m})^{- 2} \right],
\end{equation}
\end{subequations}
and using again (\ref{eq3}), as well as the Hamiltonian equations $\dot {z}_{n} =
p_{n} $ entailed by (\ref{eq5}), it is immediately seen, via (\ref{eq5}), that the
right-hand side of (\ref{eq25}) coincides with the right-hand side of (\ref{eq18}).
But the left-hand sides obviously coincide as well, so the result is proven.
(We provided here a proof of this result for completeness,
although of course this result can as well be directly inferred via the trick
from the analogous finding recently proved for the system (2.5)~[7]).

\section{The analytic structure of the solutions of (\ref{eq11})}

In this section we discuss the analytic structure of the solutions
$\underline {\zeta}  (\tau)$ of the system of ODEs (\ref{eq11}), considered
as functions of the complex variable $\tau $, and in particular we show that
the singularities of these functions are branch points of
\textit{square-root} type (see below).

These singularities are of course associated with values of the dependent
variables~$\zeta _{n} $ that cause the right-hand side of (\ref{eq11}) to diverge,
namely they are associated with ``collisions'' of two or more of the
coordinates $\zeta _{n} $. (By definition the ``collision'' of two or more
coordinates occurs when their values coincide; but we use inverted commas to
underline that, since we are considering \textit{complex} values of the
independent variable $\tau $, such ``collisions'' need not correspond to
actual collisions of the ``particles'' the motion of which as the
\textit{real} ``time'' variable $t$ unfolds is described by the ``physical''
equations of motions (\ref{eq1}) or (\ref{eq4}) --- we shall return to this point
below). Clearly the generic case corresponds to \textit{two-body}
collisions, occurring at some value $\tau = \tau _{b} $ such that, say,
\begin{subequations}
\begin{equation}
\label{eq26}
\zeta _{1} (\tau _{b}) = \zeta _{2} (\tau _{b}).
\end{equation}
Here and below, without loss of generality, when discussing two-body
collisions, we focus on the
two particles carrying the labels $1$ and $2$. Note that it is natural to
expect that this equation, (\ref{eq26}), have one or more (possibly an infinity)
of solutions $\tau _{b} $ in the complex $\tau $-plane for any
\textit{generic} solution $\underline {\zeta} (\tau)$ of
the ODEs (\ref{eq11}); while the equations characterizing a multiple collision, say
\begin{equation}
\label{eq27}
\zeta _{1} (\tau _{b}) = \zeta _{2} (\tau _{b}) = \zeta _{3} (\tau _{b}) =\cdots = \zeta _{M}
(\tau _{b}),
\end{equation}
\end{subequations}
with $2 < M < N$ have generally no solution at all (for a \textit{generic}
solution $\underline {\zeta} (\tau)$ of the ODEs (\ref{eq11});
there exist of course \textit{special} solutions of the ODEs (\ref{eq11}) which
feature such multiple collisions, and we discuss them below to demonstrate
that they as well feature branch points of \textit{square-root} type; we
moreover believe, due to the scaling character of the equations of motion
(\ref{eq11}), that \textit{completely multiple} collisions characterized by $M = N$
are, as well as \textit{two-body} collisions, featured by any
\textit{generic} solution of (\ref{eq11})~--- as suggested by the analysis of the
three-body case, see Appendix~B).

To demonstrate that the singularity of the solution $\underline {\zeta}
(\tau)$ of the ODEs (\ref{eq11}) associated with the collision
(\ref{eq26}) is a branch point of \textit{square-root} type we introduce the
following \textit{ansatz}, valid in the neighborhood of $\tau = \tau _{b} $:
\begin{subequations}
\begin{gather}
\label{eq28}
\zeta_{1}(\tau) = b + \beta (\tau-\tau_{b})^{1/2}
+ v(\tau-\tau_{b})+a(\tau-\tau_{b})^{3/2}
+ \sum_{l = 4}^{\infty} \alpha_{l}^{(1)} (\tau-\tau_{b})^{l/2},
\\
\label{eq29}
\zeta _{2}(\tau) = b - \beta (\tau-\tau_{b})^{1/2}
+ v(\tau-\tau_{b})-a(\tau-\tau_{b})^{3/2}
+ \sum_{l = 4}^{\infty}\alpha_{l}^{(2)}(\tau-\tau_{b})^{l/2},
\\
\label{eq30}
\zeta_{n}(\tau) = b_{n} + v_{n}(\tau-\tau_{b})
+ \sum_{l = 4}^{\infty} \alpha_{l}^{(n)} (\tau-\tau_{b})^{l/2},\qquad n =
3,4,\ldots,N.
\end{gather}
\end{subequations}
The fact that this \textit{ansatz} is consistent with the two-body-collision
condition (\ref{eq26}) is plain. It can moreover be shown (see Appendix~A) that
this \textit{ansatz}, (3.2), is compatible with the ODEs (\ref{eq11}), and moreover
that, while the coefficients $\beta $ and $\alpha _{l}^{(n)}
$ featured by it are uniquely determined by the requirement that
(3.2) satisfy (\ref{eq11}), the remaining coefficients, namely the $4$ constants
$b$, $v$, $a$ and $\tau _{b} $, see (\ref{eq28}), (\ref{eq29}), and the $2(N - 2)$
constants $b_{n}$, $v_{n} $, see (\ref{eq30}), are \textit{arbitrary}
(except for the obvious requirements $b_{n} \ne b$,
$b_{n} \ne b_{m} $)~---
and since altogether the number of these \textit{arbitrary} constants is
$2N$, we infer that this \textit{ansatz}, (3.2), is adequate to
represent in the neighborhood of $\tau=\tau_{b}$
the \textit{general solution} of the system of $N$
\textit{second-order} ODEs (\ref{eq11}).

We therefore conclude that the singularities associated with the generic
solution of (\ref{eq4}) are branch points of \textit{square-root} type, since
clearly this is the singularity exhibited by the right-hand sides of (3.2)
--- on the assumption that the infinite series they feature do indeed
converge in a sufficiently small neighborhood of $\tau = \tau _{b} $.

It is easily seen that this same conclusion obtains if consideration is
extended to include \textit{multiple} collisions, see (\ref{eq27}). Indeed the
corresponding \textit{ansatz} reads, in analogy to (3.2),
\begin{subequations}
\begin{gather}
\zeta _{n} (\tau) = b + \beta _{n} (\tau - \tau_{b})^{1/2} + v(\tau - \tau _{b})\nonumber\\
\phantom{\zeta _{n} (\tau) =}{} + a_{n}
(\tau - \tau _{b})^{3/2} + \sum\limits_{l = 4}^{\infty}
 \alpha _{l}^{(n)} (\tau - \tau _{b})^{l/2},
\qquad n = 1,2,\ldots,M,\label{eq31}
\\
\label{eq32}
\zeta _{n} (\tau) = b_{n} + v_{n} (\tau - \tau _{b} )
+ \sum\limits_{l = 4}^{\infty}  \alpha _{l}^{(n)}
(\tau - \tau _{b})^{l/2},\qquad n = M + 1,M
+ 2,\ldots,N.
\end{gather}
\end{subequations}

The consistency of this \textit{ansatz} with the ODEs (\ref{eq11}) is also
demonstrated in Appendix~A; but note that, as shown there, only the
$3$~constants $b$, $v$, $\tau _{b} $ appearing in the right-hand side of~(\ref{eq31}),
and a common rescaling factor, say $a$, of the $M$ coefficients $a_{n}$,
as well as the constants $b_{n}$,
$v_{n} $ appearing in the right-hand side
of (\ref{eq32}), are now \textit{arbitrary}; so, the \textit{ansatz} (3.3) features
altogether $4+2(N-M)=2N-2(M-2)$ arbitrary constants. Hence for $M > 2$ this \textit{ansatz}
does \textit{not} feature the full complement of $2N$ arbitrary constants
required in order that (3.3) represent, for $\tau \approx \tau _{b} $, the
\textit{general solution} of (\ref{eq11})~--- as indeed we expected (since we do
\textit{not} expect the \textit{generic} solution of (\ref{eq11}) to feature
\textit{multiple} collisions, at least with $2 < M < N$).

Let us end this section by reporting the \textit{similarity solution} of
(\ref{eq11}) that indeed features an \textit{$N$-body collision}. It reads, as can be
easily verified,
\begin{subequations}
\begin{equation}
\label{eq33}
\zeta _{n} (\tau) = b + \beta _{n} (\tau - \tau _{b})^{1/2} + v(\tau - \tau _{b}),
\end{equation}
with the $3$ constants $b$, $v$ and $\tau _{b} $ arbitrary, while the $N$
constants $\beta _{n} $ are instead determined by the $N$ algebraic
equations
\begin{equation}
\label{eq34}
\beta _{n} = - 4\sum\limits_{m = 1,\; m \ne n}^{N} g_{nm} (\beta
_{n} - \beta _{m})^{ - 3}.
\end{equation}
\end{subequations}

As it is well known (see for instance [2]), in the \textit{integrable}
equal-coupling-constants case,
\begin{subequations}
\begin{equation}
\label{eq35}
g_{nm} = g,
\end{equation}
the solution of (\ref{eq34}) is
\begin{equation}
\label{eq36}
\beta _{n} = ( - 2g)^{1/4}\xi _{n},
\end{equation}
where the $N$ real numbers $\xi _{n} $ are the $N$ zeros of the Hermite
polynomial $H_{N} (\xi)$ of degree~$N$,
\begin{equation}
\label{eq37}
H_{N} (\xi _{n}) = 0.
\end{equation}
\end{subequations}
Note that the formula (\ref{eq36}) defines in fact $4$ different
sets of coefficients $\beta_{n}$, due to the $4$~possible
determinations of the fourth root appearing in the right-hand side
of this equation.

Finally let us emphasize that these findings entail that, even in the
\textit{integrable} equal-coupling-constants case, see (\ref{eq35}), the solutions
of the ODEs (\ref{eq11}) do \textit{not} possess the so-called Painlev\'e property,
namely they are \textit{not} free of \textit{movable} branch points, indeed
their branch points (of \textit{square-root} type, see for instance the
special solution (\ref{eq33})) occur at values $\tau = \tau _{b} $ of the
independent variable $\tau $ which are not \textit{a priori} predictable,
but rather depend, in the context of the initial-value problem, on the
initial data. But in the \textit{integrable} case the number of these branch
points is always finite: they are indeed determined by the requirement that
the monic polynomial of degree $N$ in the variable, say, $\zeta $, the $N$ zeros
of which are the $N$ coordinates $\zeta _{n} (\tau)$ (and the
coefficients of which are themselves polynomials of degree $N$ in $\tau $),
have a \textit{double zero} (see for instance~[2]).

\section{Periodicity of the solutions of the system of ODEs (\ref{eq4}): \\
theoretical considerations}

In this section we analyze the implications of the findings of the previous
Section 3 as regards the periodicity of the solutions of the ODEs (\ref{eq4}).

We know of course that, if a solution $\underline {\zeta}  (\tau)$
of (\ref{eq11}) is holomorphic as a function of the complex
variable $\tau $ \textit{inside} the circle $\tilde {C}$ (see
Section~2)~--- and we know that such solutions do exist, in fact
in the context of the initial-value problem they emerge out of a
set of initial data which has \textit{infinite} measure in the
phase space of such data~[1]~--- then the solution~$\underline {z}
(t)$ of (\ref{eq4}) that corresponds to it via (2.4) is
\textit{completely periodic} with period~$T$, see~(\ref{eq6}),
\begin{subequations}
\begin{equation}
\label{eq38}
z_{n} (t + T) = z_{n} (t).
\end{equation}
But the transformation (2.4) actually implies, as already noted in Section~2,
an additional information, namely that in this case $\underline {z}
(t)$ is \textit{completely antiperiodic} with period $T/2$,
\begin{equation}
\label{eq39}
z_{n} (t + T/2) = - z_{n} (t).
\end{equation}
\end{subequations}
Note that (\ref{eq39}) implies (\ref{eq38}), while of course (\ref{eq38}) does not imply
(\ref{eq39}).

Let us instead assume that a branch point of a solution
$\underline {\zeta} (\tau)$ of (\ref{eq11}), occurring, say, at
$\tau = \tau _{b} $, does fall \textit{inside} the circle $\tilde
{C}$ in the complex $\tau $-plane (see Section~2). Then, due to
the \textit{square-root} nature of this branch point, see (3.2),
the evolution of the solution~$\underline {\zeta} (\tau)$ of
(\ref{eq11}) as the \textit{real} time variable~$t$ unfolds is
obtained by following the \textit{complex} time-like variable
$\tau $ as it travels (2.4) along the circular contour $\tilde
{C}$ on a \textit{two-sheeted} Riemann surface. Clearly the change
of variable (\ref{eq9}), (\ref{eq10}) entails then that the
corresponding solution~$\underline {z} (t)$ of (\ref{eq4}) is just
as well \textit{completely periodic} with period $T$, see
(\ref{eq38}), although in this case (\ref{eq39}) does no more
hold. And of course this conclusion holds provided \textit{only
one} branch point of the solution $\underline {\zeta}  (\tau)$ of
(\ref{eq11}) falls inside the circle $\tilde {C}$ in the main
sheet of the Riemann surface associated with this solution, and
\textit{no other} branch point occurs inside the circle $\tilde
{C}$ in the second sheet of this Riemann surface, namely on the
sheet entered through the cut associated with the branch point
occurring inside $\tilde {C}$ on the main sheet of the Riemann
surface (of course this Riemann surface might feature many other
sheets associated with other branch points occurring elsewhere
hence not relevant to our present discussion).

Let us now continue this analysis by considering, more generally, a solution
$\underline {\zeta}  (\tau)$ of~(\ref{eq11}) that possibly contains
more than one branch point inside the circle $\tilde {C}$ in the
main sheet of its Riemann
surface (that do not cancel each other) so that by traveling along the circle
$\tilde {C}$ \textit{several} additional Riemann sheets are accessed from
the main sheet, and let us moreover assume that, on these additional sheets,
\textit{additional} branch points possibly occur inside the circle $\tilde
{C}$ which give access to other sheets, and that possibly on these other
sheets there be \textit{additional} branch points and so on. Let in
conclusion $B$ be the \textit{total number} of \textit{additional} sheets
accessed by a point traveling around and around on the circle $\tilde {C}$
on the Riemann surface associated with the solution $\underline {\zeta}
(\tau)$ of (\ref{eq11}). This number $B$ might coincide with the
total number of branch points occurring, inside the circle~$\tilde {C}$, on
this Riemann surface~--- on all its sheets~--- or it might be smaller. Indeed,
since each of these branch points is of \textit{square-root} type, each of
the associated cuts~--- if entered into~--- gives access to \textit{one
additional} sheet. But not all these sheets need be accessed; the total
number $B$ that are actually accessed depends on the structure of the
Riemann surface, for instance no \textit{additional} sheet at all is
accessed if there is no branch point on the \textit{main} sheet of the
Riemann surface~--- even though other branch points may be present inside the
circle~$\tilde {C}$ on other sheets of the Riemann surface associated with
the solution $\underline {\zeta}  (\tau)$ of (\ref{eq11}). (It
might also be possible that different branch points cancel each other
pairwise as is the case for two branch points that are on the same sheet inside the circle
$\tilde {C}$ and generate a cut that starts at one of them and ends at the other).
In any case the overall time requested for the point $\tau
(t)$ traveling on the Riemann surface to return to its point
of departure (say, $\tau (0) = 0$ on the \textit{main} sheet)
is $(B + 1)T/2$, since a half-period $T/2$, see (\ref{eq10}), is
required to complete a tour around the circle $\tilde {C}$ on each sheet,
and the number of sheets to be traveled before getting back to the point of
departure is overall $B + 1$ (including the \textit{main} sheet). Hence in
this case, as the real time $t$ evolves, the solution $\underline {\zeta}
(\tau)$ of (\ref{eq11}) will be \textit{completely periodic} with
period $(B + 1)T/2$. Hence (see (2.4)) if $B$ is
\textit{even} the corresponding solution $\underline {z} (t)$
of (\ref{eq4}) will be \textit{completely antiperiodic} with the same period
$(B + 1)T/2$, $\underline {z} [t + (B + 1)T/2 ] = - \underline {z} (t)$, and
\textit{completely periodic} with the ``odd'' period $(B + 1)T$,
$\underline {z} [t + (B + 1)T] =
\underline {z} (t)$. If instead $B$ is \textit{odd}, the
solution $\underline {\zeta}  (\tau)$ of (\ref{eq11}) as well as
the corresponding solution $\underline {z} (t)$ of (\ref{eq4})
will both be \textit{completely periodic} in~$t$ with the period $(B + 1)T/2$
(which might be ``even'' or ``odd''~--- of course, as an
integer multiple of the basic period $T$), $\underline {z} [t +
(B + 1)T/2] = \underline {z} (t)$ (so,
in this case, the trajectories of $\underline {z} (t)$ will
display no symmetry, in contrast to the previous case).

In this analysis the assumption was implicitly understood that the total
number $B$ of \textit{additional} sheets accessed by traveling around and
around on the circle $\tilde {C}$ on the Riemann surface associated with the
solution $\underline {\zeta} (\tau)$ of (\ref{eq11}) be
\textit{finite} (a number $B$ which, as we just explained, might coincide
with, or be smaller than, the total number of branch points of that Riemann
surface that are located \textit{inside} the circle $\tilde {C}$ in the
complex $\tau $-plane); and moreover we implicitly assumed that no branch
point occur exactly on the circle $\tilde {C}$. Let us now elaborate on
these two points.

If a branch point $\tau _{b} $ occurs exactly on the circle $\tilde {C}$,
then the ``physical'' equations of motion (\ref{eq4}) become singular, due to a
particle collision occurring at the \textit{real} time $t_{c} $ defined
$\mbox{mod}\left( {T/2} \right)$ (see (\ref{eq6})) by the formula
\begin{equation}
\label{eq40}
\tau _{b} = [\exp(2i\omega t_{c}) - 1]/(2i\omega).
\end{equation}
Indeed it is easy to check via (2.4) that the condition that
$t_{c} $ be \textit{real} coincides with the requirement that the
corresponding value of $\tau _{b} $, as given by (\ref{eq40}),
fall just on the circular contour $\tilde {C}$ in the complex
$\tau $-plane. The singularity is of course due to the divergence,
at the collision time $t = t_{c} $, of the right-hand side of the
equations of motion (\ref{eq4}); there is however no corresponding
divergence of the solution $\underline {z} (t)$, which rather has
a branch point of \textit{square root} type at $t = t_{c} $, see
(3.2). But of course this entails that the speeds of the colliding
particles diverge at the collision time $t=t_{c}$ proportionally
to $|t-t_{c}|^{-1/2}$, and their accelerations diverge
proportionally to $|t-t_{c}|^{-3/2}$.

There is no \textit{a priori} guarantee that the number of branch
points inside $\tilde {C}$ of a solution~$\underline {\zeta}
(\tau)$ of (\ref{eq11}) be \textit{finite}, nor that the number
$B$ of \textit{additional} sheets accessed according to the
mechanism described above by moving around the circle $\tilde {C}$
on the Riemann surface associated with that solution $\underline
{\zeta} (\tau)$ of (\ref{eq11}) be \textit{finite} (of course $B$
might be \textit{infinite} only if the number of branch points
inside $\tilde {C}$ is itself \textit{infinite}). Obviously in
such a case $(B=\infty)$, although the complex number $\tau \equiv
\tau (t)$, see (\ref{eq10}), considered as a function of the
\textit{real} ``time'' variable $t$, is still periodic with period
$T/2$, neither the solution $\underline {\zeta}  (\tau)$
of~(\ref{eq11}), nor the corresponding solution $\underline {z}
(t)$ of (\ref{eq4}), will be periodic. The question that might
then be raised is whether such a solution --- in particular, such
a solution $\underline {z} (t)$ of the ``physical'' Newtonian
equations of motion (\ref{eq1}) corresponding to the many-body
problem characterized by the Hamiltonian (1.2)~--- displays a
``chaotic'' behavior, namely, in the context of the initial-value
problem, a ``sensitive dependence'' on the initial data. We shall
return to this question below.

So far we have discussed the relation among the analytic structure
of a solution $\underline {\zeta}  (\tau)$ of~(\ref{eq11}) and the
corresponding solution $\underline {z} (t)$ of (\ref{eq4}). Let us
now return to the simpler cases considered at the very beginning
of this analysis and let us consider how the \textit{transition}
from one of the two regimes described there to the other occurs in
the context of the initial-value problem for (\ref{eq4}), and
correspondingly for (\ref{eq11}), see~(2.6). Hence let us assume
again that the initial data for (\ref{eq4}), and correspondingly
for (\ref{eq11}) (see (2.6)), entail that no branch point of the
corresponding solution~$\underline {\zeta} (\tau)$ of (\ref{eq11})
occurs \textit{inside} the circular contour $\tilde {C}$ on the
\textit{main} sheet of the associated Riemann surface, so that the
corresponding solution~$\underline {z} (t)$ of (\ref{eq4})
satisfies both (\ref{eq38}) and (\ref{eq39}). Let us imagine then
to modify with continuity the initial data, for instance by
letting them depend on an appropriate scaling parameter (a
particular way to do so will be introduced in the following
Section~5, as a~convenient technique to present numerical
results). As a consequence the branch points of the
solution~$\underline {\zeta}  (\tau)$ of (\ref{eq11}) move, and
the Riemann surface associated to this solution~$\underline
{\zeta}  (\tau)$ of (\ref{eq11}) gets accordingly modified. We are
interested in a movement of the branch points which takes the
closest one of them on the \textit{main} sheet of the Riemann
surface from outside to inside the circle~$\tilde {C}$. In the
process that branch point will cross the circle~$\tilde {C}$, and
the particular set of initial data that correspond to this
happening is then just a set of initial data that entails the
occurrence of a collision in the time evolution of the many-body
problem (\ref{eq4}), occurring at a \textit{real} time $t = t_{c}
$ defined by (\ref{eq40}), as discussed above. After the branch
point has crossed the contour $\tilde {C}$ and has thereby entered
inside the circular disk $C$, the corresponding solution
$\underline {z} (t)$ of (\ref{eq4}) is again collision-free but
its periodicity properties are changed. One might expect that the
new solution continue then to satisfy~(\ref{eq38}) but cease to
satisfy (\ref{eq39}). This is indeed a possibility, but it is not
the only one. Indeed, since the time evolution of the solution
$\underline {z} (t)$ of (\ref{eq4}) obtains via (2.4) by following
the time evolution of the corresponding solution $\underline
{\zeta}  (\tau)$ of (\ref{eq11}) as the point $\tau \equiv \tau
(t)$ goes round the circle $\tilde {C}$ on the Riemann surface
associated with that solution, the occurrence of a branch point
\textit{inside} the circle $\tilde {C}$ on the \textit{main} sheet
of that Riemann surface entails that the access is now open to a
second sheet, and then possibly to other sheets if, on that second
sheet, there also are branch points inside the circle~$\tilde
{C}$. If this latter possibility does not occur, namely if on that
second sheet there are no branch points inside the circle $\tilde
{C}$, then indeed there occurs for the corresponding solution
$\underline {z} (t)$ of~(\ref{eq4}) a transition from a
periodicity property characterized by the validity of
both~(\ref{eq38}) and~(\ref{eq39}), to one characterized by the
validity of (\ref{eq38}) but not of (\ref{eq39}). If instead there
is a least one branch point in the second sheet inside the circle
$\tilde {C}$, then the periodicity --- if any~--- featured after
the transition by the solution $\underline {z} (t)$ of (\ref{eq4})
depends, as discussed above, on the number $B$ of sheets that are
sequentially accessed before returning~--- if ever~--- to the
\textit{main} sheet.

To simplify our presentation we have discussion above the
\textit{transition} process by taking as point of departure for
the analysis the \textit{basic periodic solution}~--- that
characterized by the validity of both (\ref{eq38}) and
(\ref{eq39}), the existence of which has been demonstrated
in~[1]~--- and by discussing how a continuous modification of the
initial data may cause a~transition to a different regime of
periodicity, with the transition occurring in correspondence to
the special set of initial data that yields a solution
characterized by a particle collision, namely a set of initial
data for which the Newtonian equations of motion become singular
at a finite \textit{real} time $t_{c}$ (defined
$\mbox{mod}(T/2)$). But it is clear
 that exactly the same mechanism accounts for every
transition that occurs from a solution $\underline {z}(t)$
of (\ref{eq4}) characterized by a type of periodicity to a solution $\underline
{z} (t)$ of (\ref{eq4}) characterized by a different periodicity
regime~--- or by a lack of periodicity.

The final point to be discussed is the question we postponed above, namely
the character of the \textit{nonperiodic} solutions $\underline {z} (t)$
of (\ref{eq4}) (if any), which we now understand to be
characterized, in the context of the mechanism described above, by access to
an endless sequence of different sheets~--- all of them generated by branch
points of \textit{square-root} type~--- of the Riemann surface associated
with the corresponding solution $\underline {\zeta} (\tau)$
of~(\ref{eq11}). The following two possibilities can be imagined in this connection~---
which of course does not entail they are indeed both realized.

The first possibility --- which we surmise to be the most likely
one to be actually realized --- is that an \textit{infinity} of
such relevant branch points occur quite closely to the circular
contour $\tilde {C}$, hence that there be some of them that occur
arbitrarily close to $\tilde {C}$. This then entails that the
corresponding \textit{nonperiodic} solutions $\underline {z} (t)$
of~(\ref{eq4}) manifest a~\textit{sensitive dependence} on their
initial data (which we consider to be the signature of
a~\textit{chaotic} behavior). Indeed a modification, however
small, of such initial data entails a modification of the pattern
of such branch points, which shall cause some of them to cross
over from one to the other side of the circular contour $\tilde
{C}$. But then the two solutions $\underline {z} (t)$ of
(\ref{eq4}) corresponding to these two assignments of initial data
--- before and after the modification, however close these data
are to each other --- will eventually evolve \textit{quite
differently}, since their time evolutions are determined by access
to two \textit{different} sequences of sheets of the Riemann
surfaces associated with the two corresponding solutions
$\underline {\zeta}  (\tau)$ of (\ref{eq11})~--- two Riemann
surfaces which themselves need not be very different (to the
extent one can make such statements when comparing two objects
having as complicated a structure as a Riemann surface with an
\textit{infinite} number of sheets produced by an
\textit{infinite} number of branch points of \textit{square-root}
type). So this is the mechanism whereby a \textit{chaotic}
behavior may develop for the system (\ref{eq1})~--- but of course
not in the \textit{integrable} case with equal coupling constants
(in that case, as mentioned above, the number of branch points of
the solutions $\underline {\zeta} (\tau)$ of (\ref{eq11}) is
always \textit{finite}, since the $N$ coordinates $\zeta _{n}
(\tau)$ are in this case the $N$ zeros of a polynomial of degree
$N$ the coefficients of which are polynomials in the variable
$\tau $ [2]). Note that the emergence of such a chaotic behavior
would not be associated with a local exponential divergence of
trajectories in phase space --- it would be rather analogous to
the mechanism that causes a chaotic behavior in the case of, say,
a triangular billiard with angles which are irrational fractions
of $\pi$~--- a chaotic behavior also not due to a local separation
of trajectories in phase space, but rather to the eventual
emergence of a different pattern of reflections (indeed of any two
such billiard trajectories, however close their initial data, one
shall eventually miss a reflection near a corner which the other
one does take, and from that moment their time evolutions become
quite different).

A different possibility, which we consider unlikely but we cannot \textit{a
priori} exclude at this stage of our analysis, is that \textit{nonperiodic}
solutions $\underline {z}(t)$ of (\ref{eq4}) exist which are
associated with a Riemann surface of the corresponding solutions $\underline
{\zeta}  (\tau)$ of (\ref{eq11}) that, even though it possesses an
\textit{infinite} number of relevant branch points inside the circular
contour $\tilde {C}$, it features all of them~--- or at least most of them,
except possibly for a \textit{finite} number of them~--- located in a region
well inside $\tilde {C}$, namely separated from it by an annulus of
\textit{finite} thickness. Clearly in such a case two \textit{nonperiodic}
solutions $\underline {z} (t)$ of (\ref{eq4}) which emerge from
sufficiently close initial data separate slowly and gradually throughout
their time evolution, hence they do \textit{not} display a \textit{sensitive
dependence} on their initial data~--- hence, in such a case there would be
solutions which are \textit{nonperiodic} (nor, of course, multiply periodic)
but which nevertheless do not display a \textit{chaotic} behavior~--- or, to
be more precise, there would be sets of initial data, having nonvanishing
measure in the phase space of initial data, which yield such
\textit{nonperiodic} (yet \textit{nonchaotic}) solutions.

As indicated by its title, this paper is mainly focussed on the
\textit{periodic} solutions. In the following Section~5 the analysis of
their phenomenology given in this section is complemented by the display of
numerical simulations. We shall also exhibit solutions which appear
\textit{nonperiodic} and perhaps \textit{chaotic}, although such
characteristics can of course never be demonstrated with complete cogency
via numerical examples.

\section{Periodicity of the solutions of the system of ODEs (\ref{eq4}):
numerical simulations}

In this section we display several numerical solutions of the Newtonian
equations of motion (\ref{eq4}) with
\begin{subequations}
\begin{equation}
\label{eq41}
\omega = 2\pi
\end{equation}
hence (see (\ref{eq6}))
\begin{equation}
\label{eq42}
T = 1.
\end{equation}
\end{subequations}

These results confirm graphically the findings discussed in the preceding
Section~4. The strategy of our presentation is to exhibit a sequence of
numerically-computed solutions of (\ref{eq4}) with (\ref{eq41}) corresponding to
different choices of initial data, for various models characterized by an
assignment of the number $N$ of particles and of the values of the coupling constants $g_{nm}$
(that are always assumed to satisfy the symmetry property (\ref{eq3}), $g_{nm}=g_{mn}$).
For obvious reasons of simplicity, see below, we restrict consideration to $N=3$ and $N=4$.
For each model we consider sequences of motions characterized by sets of initial data linked
to each other by the formulas
\begin{subequations}
\begin{gather}
\label{eq43}
x_{n} (0) = \lambda ^{- 1} x_{n}^{(0)},
\qquad y_{n} (0) = \lambda ^{ - 1}y_{n}^{(0)},\\
\label{eq44}
\dot {x}_{n} (0) = \lambda \dot {x}_{n}^{(0)} - 2\pi (\lambda - \lambda ^{- 1})
y_{n}^{(0)} ,\qquad
\dot {y}_{n} (0) = \lambda \dot {y}_{n}^{(0)} + 2\pi (\lambda - \lambda ^{ - 1})
x_{n}^{(0)},
\end{gather}
of course with $z_{n} \equiv x_{n} + i y_{n} $. Here $\lambda $ is a
\textit{positive} rescaling parameter the different values of which identify
different sets of initial data (while the data $x_{n}^{(0)}$,
$y_{n}^{(0)}$, $\dot {x}_{n}^{(0)}$,
$\dot {y}_{n}^{(0)}$ are kept fixed). The motivation for such a
choice is that, as it can be easily verified by rescaling appropriately the
dependent and independent variables (see (2.3) and (2.6) with (\ref{eq41})),
these sets of initial data (5.2) identify different solutions $\underline {z} (t)$
of the system of ODEs (\ref{eq4}) with (\ref{eq41}) that correspond to
different solutions $\underline {\zeta}  (\tau)$ of the
system of ODEs (\ref{eq11}) related to each other by the following change
of initial data,
\begin{equation}
\label{eq52c}
\underline{\zeta}(0)=\lambda^{-1}\underline{\zeta}^{(0)},
\qquad \underline{\zeta}'(0)=\lambda\underline{\zeta}'{}^{(0)}.
\end{equation}
\end{subequations}
This, as can be easily verified, entails these solutions $\underline{\zeta}(\tau)$
are related to each other merely via a rescaling
of dependent and independent variables by a constant factor.
Hence all these solutions are associated
to the \textit{same} Riemann surface except for a
shrinking of the complex $\tau $-plane by a common factor --- which,
as can be readily verified, is just $\lambda ^{2}$. Therefore, in the context
of the discussion of the preceding Section~4, to analyze the motions yielded
by the initial data (5.2) one can just imagine to multiply the diameter of the
circle $\tilde {C}$ by the factor $\lambda ^{2}$ without modifying
the Riemann surface~--- so that larger values of $\lambda $ entail that more
branch points fall within $\tilde {C}$.

So for small enough values of $\lambda$~--- namely for initial
conditions characterized by large particle coordinates hence by
large interparticle separation and by large initial velocities
almost ``orthogonal'' to the initial positions (see (5.2a,b))~--- the
circle $\tilde{C}$ shrinks to a small enough radius so that it
contains no branch points, hence the corresponding motion
(solution of (\ref{eq4}) with (5.2a,b)) is \textit{completely
periodic} with period $T=1$, see (\ref{eq38}) and~(5.1b),
and moreover it features the symmetry property (\ref{eq39}). As
$\lambda$ gets increased one goes through the scenarios described
in Section~4~--- since, as we concluded above, such an increase can
be interpreted as amounting merely to an increase of the radius of
the circle~$\tilde{C}$, causing thereby more and more branch
points to be enclosed by it. For very large~$\lambda$ this process
can cause \textit{all} branch points to be enclosed
\textit{inside}~$\tilde{C}$, a situation which is of course
equivalent, as regards the periodicity of the solutions of
(\ref{eq1}) as functions of the \textit{real} time variable $t$,
to all of them being \textit{outside}~--- hence the expectation,
that for very large $\lambda$ the motion be again
\textit{completely periodic} with period $T=1$ and also
symmetrical, see (\ref{eq38}), (\ref{eq39}) (but this need not
necessarily happen, since the possibility that the solution
$\underline{\zeta}(\tau)$ possesses branch points at infinity
cannot be excluded). This expectation is indeed confirmed by some
of the examples reported below, while in other cases for very
large $\lambda$ the numerical simulations become too difficult to
be performed reliably, and the corresponding trajectories become
unsuitable for transparent display~--- note that for very large
$\lambda$ the particles are initially all very close to the origin
(hence to each other) and with large initial velocities,
see~(5.2a,b).

But in this paper we do not elaborate on the numerical aspects, except to
reassure the skeptic reader that we made sure in each case of the
reliability of the results presented below; we refer for more details to~[6].
 Let us however note that, while one might think that a useful check of
the accuracy of the computation is provided by a verification that the
numerically evaluated coordinates $z_{n}(t)$ indeed imply
that the collective coordinates $\bar {z}(t)$ respectively
$\bar {z}^{(2)} (t)$, see (\ref{eq14}) respectively
(\ref{eq17}), evolve periodically with periods $T=1$ respectively $T/2=1/2$ according to
(\ref{eq16}) respectively (\ref{eq19}), these tests are in fact not at all cogent:
indeed, even poorly evaluated coordinates $z_{n} (t)$ tend to
yield collective coordinates $\bar {z}(t)$ and $\bar
{z}^{(2)} (t)$ that evolve properly. This
happens because the equations of motion that determine the evolution of the
collective coordinates $\bar {z}(t)$ and $\bar {z}^{(2)}
(t)$ are in fact so simple (see (2.7) and (2.8)),
that numerical errors made in the integration of (\ref{eq4}) tend to cancel out
when $\bar {z}(t)$ and $\bar {z}^{(2)}(t)$ are evaluated.

Let us also note that, in all the examples considered below, we assigned for simplicity
real integer values to the coupling constants; but we did check in every case that the
qualitative character of the motions does not change if these coupling constants are
replaced by (neighboring) values which are neither entire nor real.

The first example we consider is characterized by the following parameters:
\begin{subequations}
\begin{equation}
\label{eq46}
N=3; \qquad g_{12}=g_{21}=1,\qquad  g_{23}=g_{32}=10, \qquad g_{31}=g_{13}=-2,
\end{equation}
and by the following values of the parameters $x_{n}^{(0)}$,
$y_{n}^{(0)}$, $\dot{x}_{n}^{(0)}$, $\dot{y}_{n}^{(0)}$
characterizing the initial data via (5.2):
\begin{gather}
x_{1}^{(0)}=1, \qquad y_{1}^{(0)}=0, \qquad \dot{x}_{1}^{(0)}=0, \qquad
 \dot{y}_{1}^{(0)}=-1, \nonumber\\
x_{1}^{(0)}=0, \qquad y_{1}^{(0)}=1, \qquad \dot{x}_{1}^{(0)}=-2, \qquad
 \dot{y}_{1}^{(0)}=0, \nonumber\\
x_{1}^{(0)}=0.5,\qquad y_{1}^{(0)}=0.5, \qquad
 \dot{x}_{1}^{(0)}=-1, \qquad \dot{y}_{1}^{(0)}=0.
\label{eq47}
\end{gather}
\end{subequations}

The following Table 5.1 provides an overview of the main features
(periodicity, symmetry) of the motions that emerge from initial data
determined by (5.2) with (\ref{eq43}),
as well as an indication of the figures (if any) where the corresponding
trajectories are displayed.
Here and always below the trajectories of particle $1$,
$2$ respectively $3$ are shown in red, green, respectively
blue (and below, those of particle $4$ in yellow). Whenever we felt such an additional indication
might be usefully displayed we indicated with a black diamond the \textit{initial}
position of each particle
(at $t=0$), with a black dot the position at a subsequent time $t=t_{1}$ (generally
chosen to coincide with some
fraction of the period, $t_{1}=T/p$, with $p$ a conveniently chosen \textit{positive integer}),
and with smaller black dots the position at every subsequent integer multiple of $t_{1}$
(namely at $t=t_{k}=k t_{1}$, $k=2,3,\ldots$); in this manner
the direction of the motion along the trajectories can be inferred
(from the relative positions along the
trajectories of the diamond and the larger dot), as well as some indication
of the positions of the particles
over time, as they move (by counting the dots along the trajectory). Of course
a much more satisfactory
visualization of the motions is provided by simulations in which the particle
motions are displayed as they
unfold over time (as in a movie); it is planned to make available soon, via the web,
the numerical code suitable
to perform such simulations on personal computers~[6].
Let us emphasize that such simulations are particularly
stunning to watch in the case of high-period trajectories, which are
very complicated (see below), so that the
fact that the particles return eventually \textit{exactly} on their tracks
appears quite miraculous and is indeed a
remarkable proof of the reliability of the numerical computation.

\begin{center}
\small

Table 5.1
\medskip

\begin{tabular}{|c|c|c|c|c|c|c|c|c|c|c|}
\hline
$\lambda$ & 0.5 & 0.712 & 0.81 & 0.85 & 0.9 & 0.97 & 0.974 & 0.975 & 0.98 & 0.99 \\
\hline
Period & 1 & 1 & 1 & 1 & 1 & HSL & 15 & 15 & 13 & 11 \\
\hline
Symmetry & Yes & Yes & No & No & No & --- & No & No & No & No \\
\hline
Fig. 5.1 & a & b & c & --- & d & --- & e & --- & --- & f \\
\hline
\end{tabular}

\medskip

\begin{tabular}{|c|c|c|c|c|c|c|c|c|c|c|}
\hline
$\lambda$ & 1 & 1.01 & 1.02 & 1.03 & 1.04 & 1.05 & 1.1 & 1.2 & 2 & 5 \\
\hline
Period & 9 & 9 & 7 & 5 & 5 & 3 & 1 & 1 & 1 & 1 \\
\hline
Symmetry & No & No & No & No & No & No & No & No & No & Yes \\
\hline
Fig. 5.1 & g & --- & h & i & --- & j & ---& k & --- & m \\
\hline
\end{tabular}
\end{center}

Now some comments on these results (see Table~5.1 and the set of
Figs.~5.1). For $\lambda \leq 0.712$ there clearly are no
singularities inside $\tilde{C}$, hence the motion is periodic
with the basic period $T=1$, and moreover antiperiodic with period
$T=1/2$, so that all trajectories are symmetrical (see for
instance Figs.~5.1a,b). A transition occurs at some value of
$\lambda$ larger than $0.712$ but smaller than $0.81$, and it
causes a branch point to enter inside $\tilde{C}$, opening the way
to a second sheet (of the Riemann surface associated with the
solution~$\underline{\zeta}(\tau)$ of (2.5) with~(5.3)) and
thereby causing the trajectories of the solution
$\underline{z}(t)$ of (2.1) to loose their symmetry, while still
preserving period $T=1$. This transition is due to a collision
among particles $1$ (red) and $3$ (blue), occurring at a time of
the order of, or maybe a bit less than, $(3/10)T=3/10$ (see
Figs.~5.1b and~5.1c, where clearly $t_{1}=T/10=1/10$). Another,
more dramatic, transition occurs for a value of $\lambda$
between~$0.9$ and $0.97$, and it causes a major increase in the
complication of the motion, perhaps a transition to chaos. This
transition is of course due to the entrance inside $\tilde{C}$ of
a branch point that opens the way~--- as the complex variable
$\tau$ travels around and around on the circle $\tilde{C}$~--- to
a very large number of sheets of the Riemann surface associated
with the solution $\underline{\zeta}(\tau)$ of (2.5) with~(5.3),
possibly to an infinity of them, resulting in motions that look
chaotic: this we indicate in Table~5.1 with the acronym HSL, which
stands for ``Hic Sunt Leones''~--- see the column with
$\lambda=0.97$. This transition is due again to a collision
between particles $1$ (red) and $3$ (blue), at a time of the
order, or maybe a little less, than $(8/10)T=8/10$ (see Fig.~5.1d,
where clearly again $t_{1}=T/10=1/10$). Further increases of
$\lambda$ cause instead a \textit{decrease} of the complication of
the trajectories, characterized first of all by a return to
periodic (if still very complicated, due to the large periods)
motions, and subsequently by a progressive decrease of the period
(see the relevant graphs, from Fig.~5.1e to Fig.~5.1m); this is of
course interpreted as due to the fact that the access to
additional sheets of the Riemann surface associated with the
solution $\underline{\zeta}(\tau)$ of (2.5) with~(5.3) gets shut
off because also the second branch point corresponding to the cut
that opened the way to those sheets gets enclosed inside the
circle $\tilde{C}$. Thus the trajectories shown in Fig.~5.1e
(corresponding to $\lambda=0.974$; see Table~5.1) are interpreted,
on the basis of the discussion of the preceding Section~4, as
corresponding to access to $30$ sheets altogether of the Riemann
surface associated with the solution $\underline{\zeta}(\tau)$ of
(2.5) with (5.3); likewise those shown in Fig.~5.1f (corresponding
to $\lambda=0.99$; see Table~5.1) are interpreted as corresponding
to access to $22$ sheets of the Riemann surface associated with
the solution $\underline{\zeta}(\tau)$ of (2.5) with (5.3); those
shown in Fig.~5.1k (corresponding to $\lambda=1.2$; see Table 5.1)
are interpreted as corresponding to access to $2$ sheets, and
finally those shown in Fig. 5.1m (corresponding to $\lambda=5$;
see Table~5.1) are interpreted as corresponding to access to just
$1$ sheet, namely just the main one.

The second example we consider is characterized by the following parameters:
\begin{subequations}
\begin{equation}
\label{eq48}
N=3 ;\qquad
 g_{12}=g_{21}=10, \qquad g_{23}=g_{32}=3,\qquad g_{31}=g_{13}=-10,
\end{equation}
and by the following values of the parameters $x_{n}^{(0)}$,
$y_{n}^{(0)}$, $\dot{x}_{n}^{(0)}$, $\dot{y}_{n}^{(0)}$
characterizing the initial data via (5.2):
\begin{gather}
x_{1}^{(0)}=1, \qquad y_{1}^{(0)}=0, \qquad
 \dot{x}_{1}^{(0)}=0, \qquad \dot{y}_{1}^{(0)}=2, \nonumber\\
x_{1}^{(0)}=0, \qquad y_{1}^{(0)}=1,\qquad
 \dot{x}_{1}^{(0)}=4 ,\qquad \dot{y}_{1}^{(0)}=0, \nonumber\\
x_{1}^{(0)}=0.5, \qquad y_{1}^{(0)}=0.5, \qquad
 \dot{x}_{1}^{(0)}=-0.5, \qquad \dot{y}_{1}^{(0)}=1.
\label{eq49}\end{gather}
\end{subequations}

\begin{center}
\small

Table 5.2
\medskip

\begin{tabular}{|c|c|c|c|c|c|c|c|c|c|c|}
\hline
$\lambda$ & 0.5 & 1 & 1.2 & 1.5 & 2 & 2.2 & 2.5 & 3 & 3.5 & 4 \\
\hline
Period & 1 & 1 & HSL & HSL & 14 & 14 & 17 & 17 & 17 & 17 \\
\hline
Symmetry & Yes & No & --- & --- & No & Yes & Yes & Yes & Yes & Yes \\
\hline
Fig. 5.2 & a & b & --- & c & d & --- & --- & --- & e & --- \\
\hline
\end{tabular}

\medskip

\begin{tabular}{|c|c|c|c|c|c|c|c|c|c|c|}
\hline
$\lambda$ & 4.05 & 4.15 & 4.2 & 4.5 & 5 & 10 & 18 & 20 & 30 & 50 \\
\hline
Period & 17 & 10 & 9 & 7 & 7 & 7 & 7 & 7 & 7 & 7 \\
\hline
Symmetry & Yes & No & No & No & No & Yes & Yes & Yes & Yes & Yes \\
\hline
Fig. 5.2 & --- & f & g & --- & h & --- & i & --- & --- & --- \\
\hline
\end{tabular}
\end{center}

Since the qualitative picture is analogous to that displayed above, we limit
our presentation to the display of Table~5.2 and of the corresponding set of
graphs (see Figs.~5.2), letting to the alert reader the fun to repeat the discussion
as given above.  We merely emphasize that, in contrast to the previous case, we see in this
case also trajectories with an \textit{even} period, as well as some high-period trajectories
that are \textit{symmetrical} (the alert reader will have no difficulty in identifying the number
of sheets of the Riemann surface associated with the solution $\underline{\zeta}(\tau)$
of (2.5) with (5.4),
for each of the cases reported in Table~5.2). One is also led to suspect, from the data
reported in the last columns of Table~5.2, that the Riemann surface associated with the
solution $\underline{\zeta}(\tau)$ of (2.5) with (5.4) possesses (at least)
$7$ branch points at or near infinity
(of course on different sheets, so that they do not cancel each other).

The third example we consider is a $4$-particle one, characterized by the following parameters:
\begin{subequations}
\begin{gather}
N=4; \qquad g_{12}=g_{21}=-3, \qquad g_{13}=g_{31}=9, \qquad
 g_{14}=g_{41}=-12, \nonumber\\
g_{23}=g_{32}=-9 ,\qquad  g_{24}=g_{42}=12, \qquad  g_{34}=g_{43}=3 ,
\label{eq50}
\end{gather}
and by the following values of the parameters $x_{n}^{(0)}$,
$y_{n}^{(0)}$, $\dot{x}_{n}^{(0)}$, $\dot{y}_{n}^{(0)}$
characterizing the initial data via (5.2):
\begin{gather}
x_{1}^{(0)}=-2.4, \qquad  y_{1}^{(0)}=0 ,\qquad
 \dot{x}_{1}^{(0)}=0, \qquad \dot{y}_{1}^{(0)}=1, \nonumber\\
x_{1}^{(0)}=2.4, \qquad y_{1}^{(0)}=0, \qquad \dot{x}_{1}^{(0)}=-2,\qquad
 \dot{y}_{1}^{(0)}=-2, \nonumber\\
x_{1}^{(0)}=0, \qquad y_{1}^{(0)}=0.5, \qquad
 \dot{x}_{1}^{(0)}=-0.5, \qquad \dot{y}_{1}^{(0)}=-2, \nonumber\\
x_{1}^{(0)}=0.5, \qquad y_{1}^{(0)}=0.5, \qquad \dot{x}_{1}^{(0)}=-0.5, \qquad
 \dot{y}_{1}^{(0)}=1.
\label{eq51}\end{gather}
\end{subequations}

\newpage

\begin{center}
\includegraphics[width=15cm]{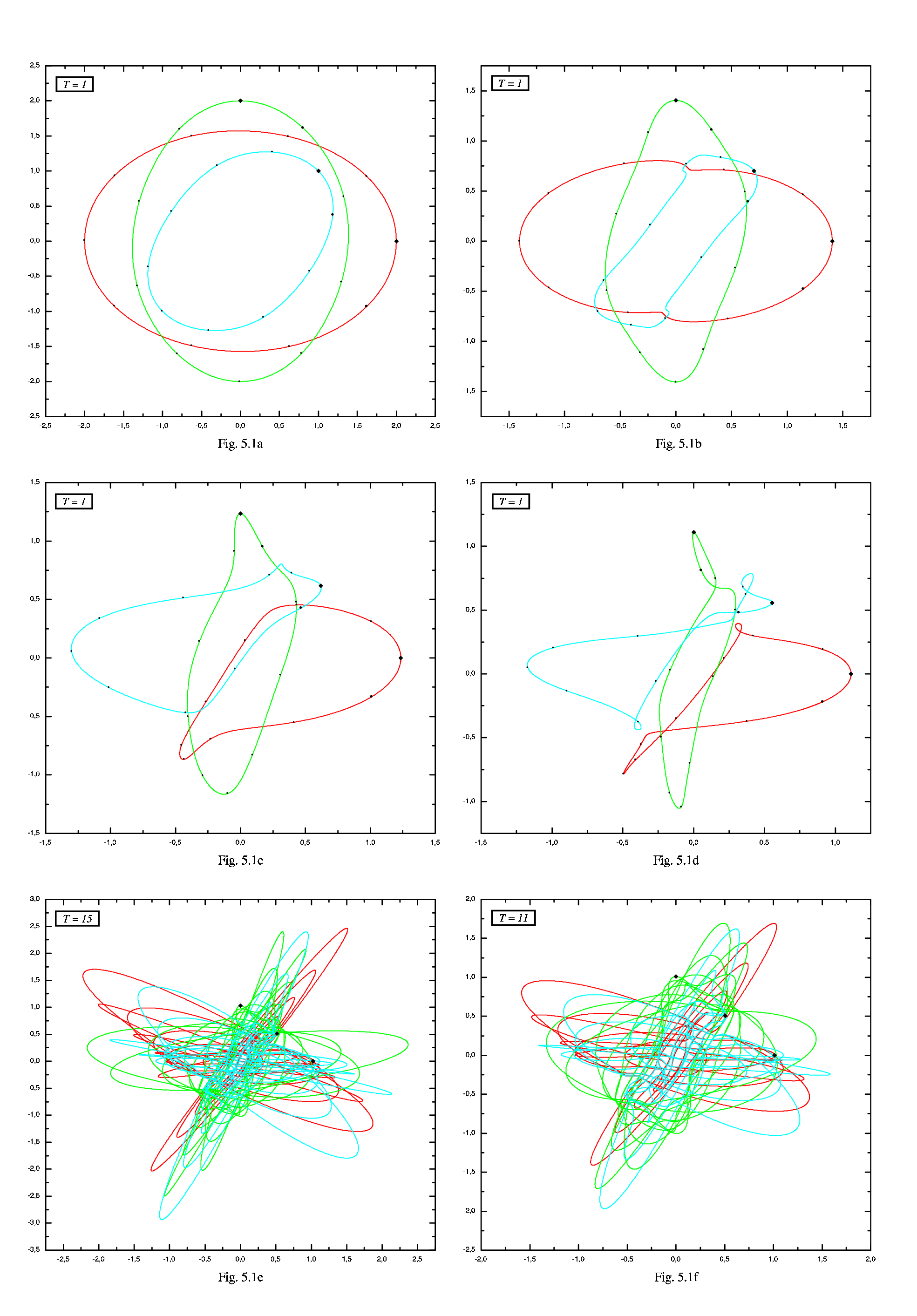}
\end{center}

\newpage

\begin {center}
\includegraphics[width=15cm]{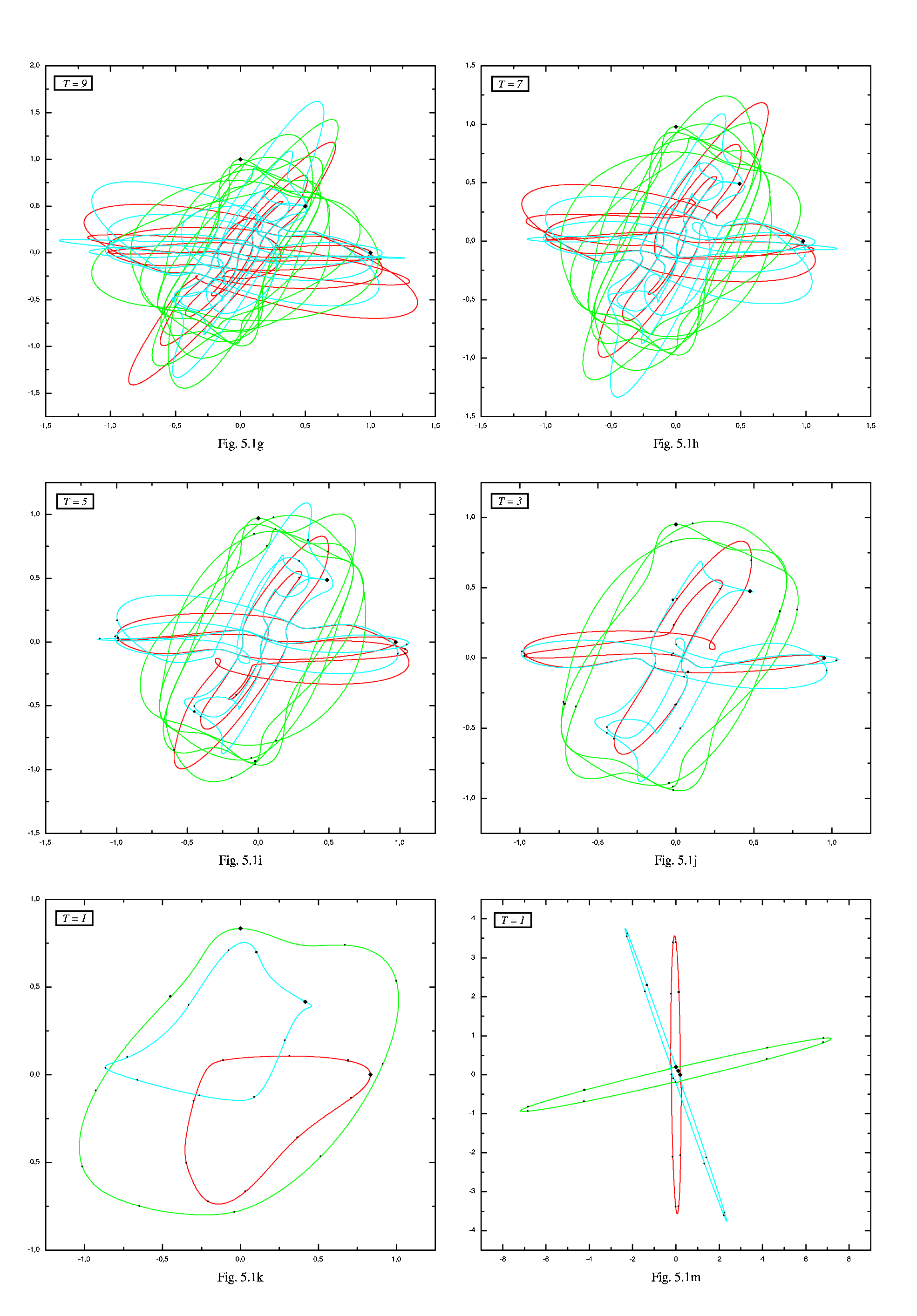}
\end {center}

\newpage

\begin {center}
\includegraphics[width=15cm] {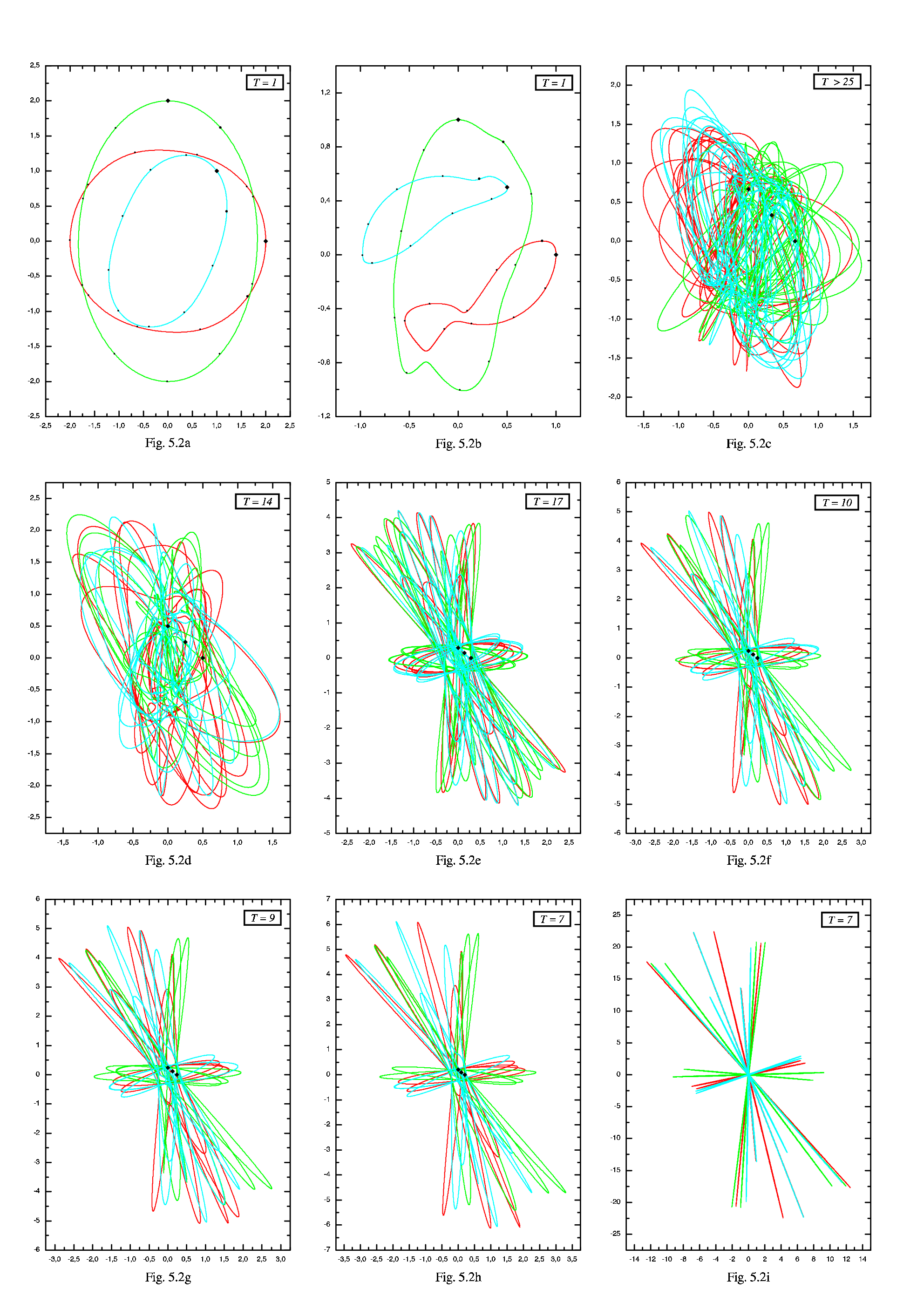}
\end {center}

\newpage

\begin {center}
\includegraphics[width=15cm] {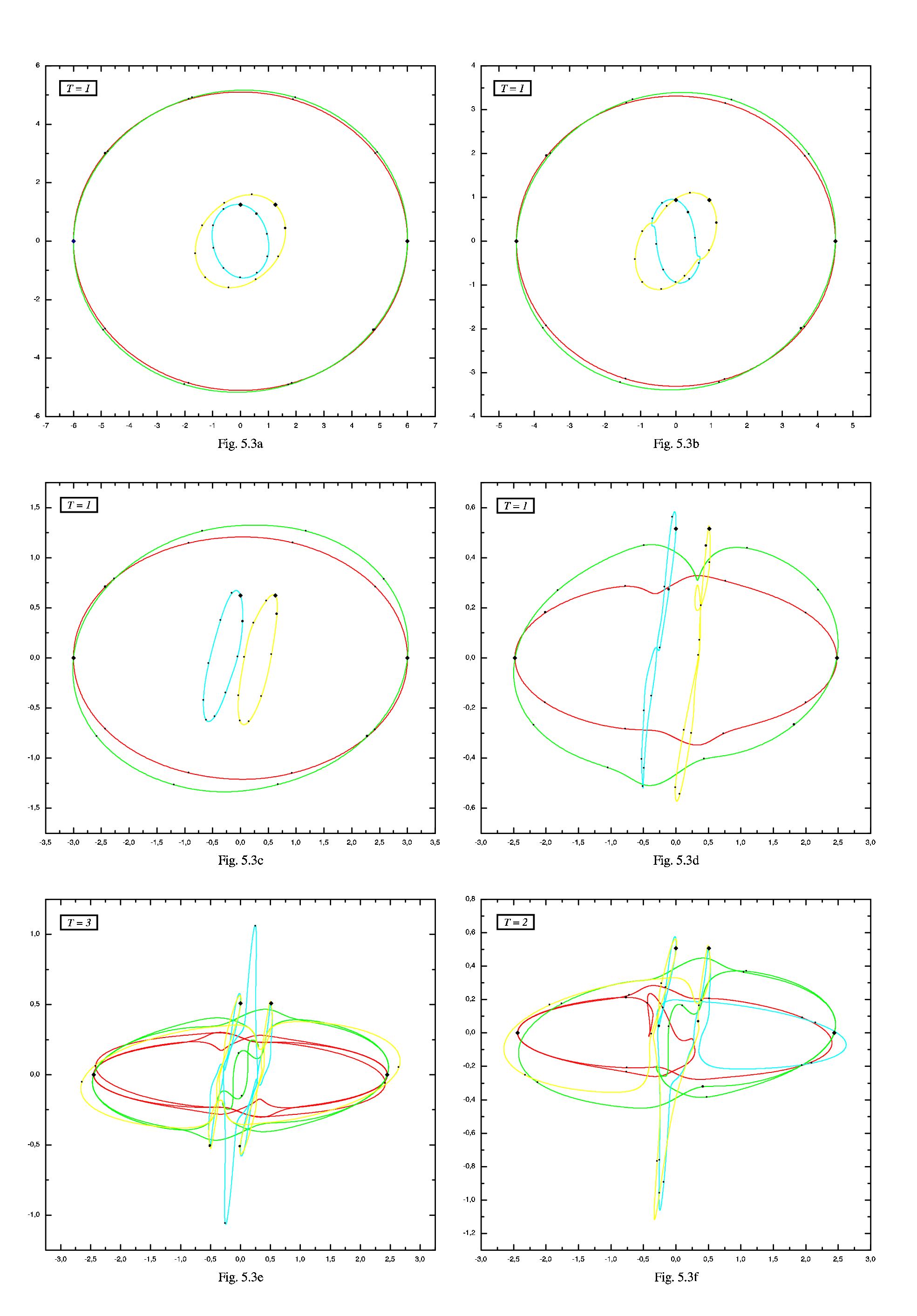}
\end {center}

\newpage

\begin {center}
\includegraphics[width=15cm] {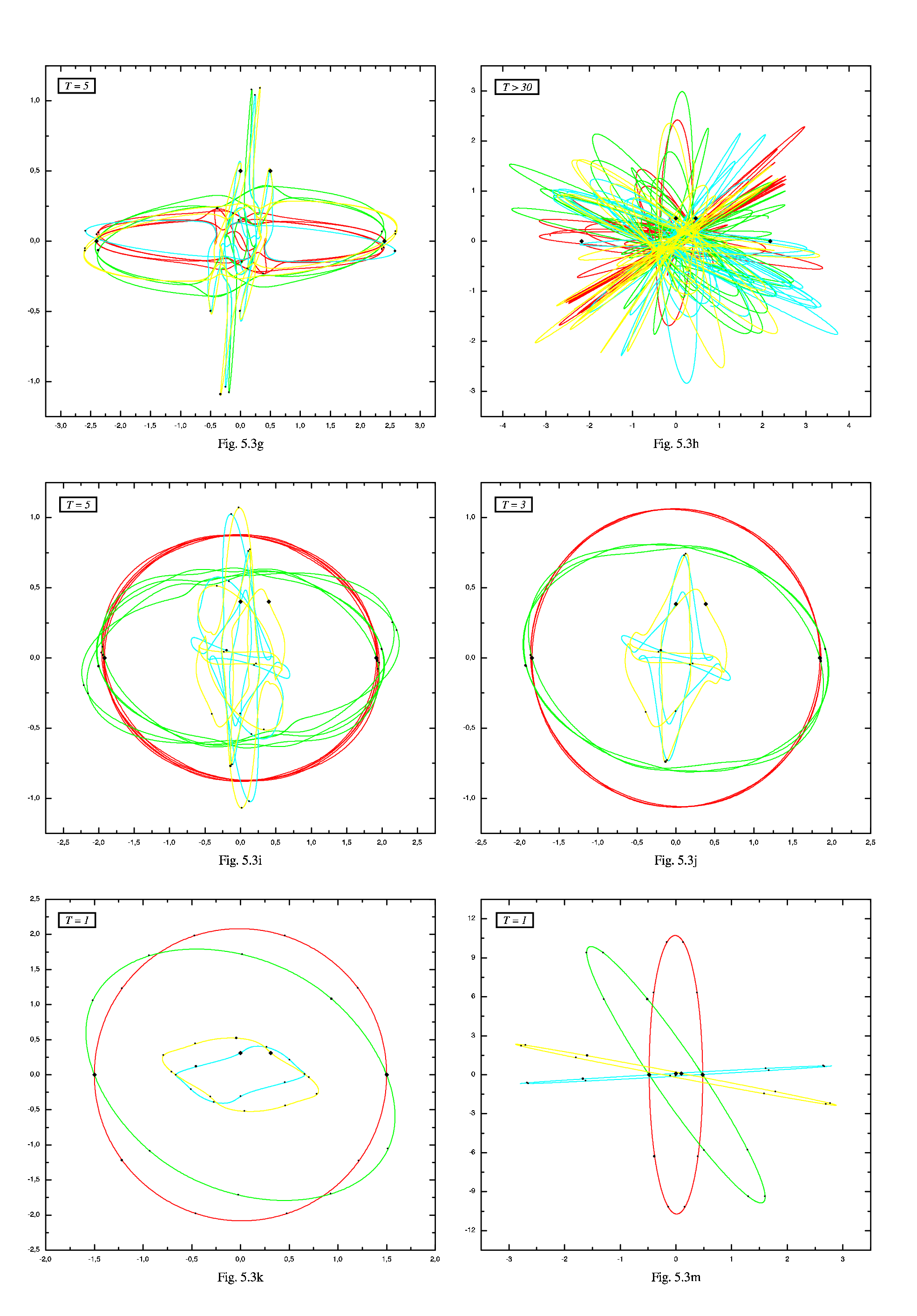}
\end {center}

\newpage

\begin {center}
\includegraphics[width=15cm] {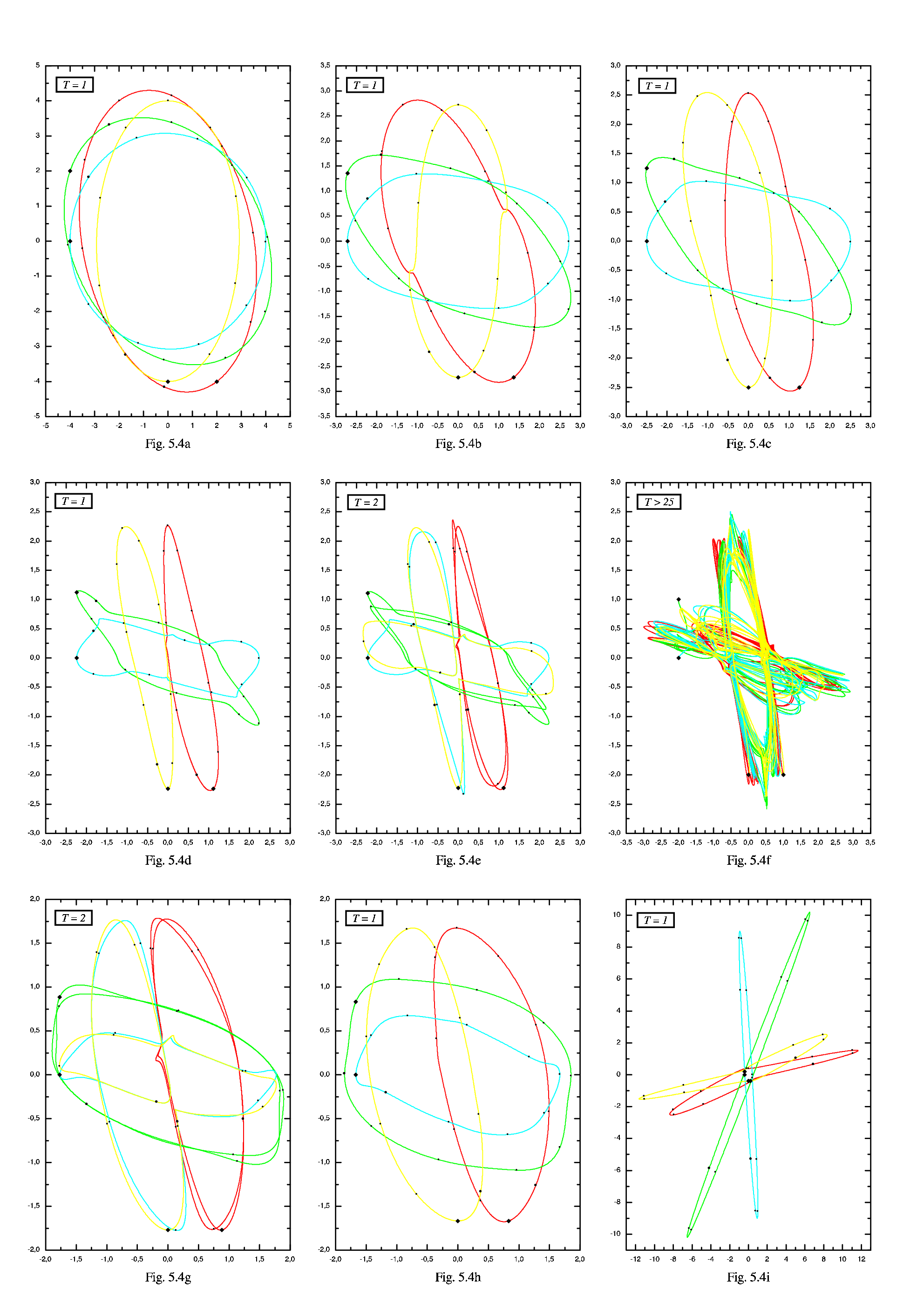}
\end {center}

\newpage

\noindent
A representative selection of our findings is reported
in Table~5.3, and some of the corresponding trajectories are
displayed in the Figures as indicated there.

\begin{center}
\small Table 5.3
\medskip

\begin{tabular}{|c|c|c|c|c|c|c|c|c|c|}
\hline
$\lambda$ & 0.4 & 0.533 & 0.8 & 0.9 & 0.9687 & 0.98 & 0.985 & 0.99 & 1 \\
\hline
Period & 1 & 1 & 1 & 1 & 1 & 3 & 2 & 5 & 5 \\
\hline
Symmetry & Yes & Yes & No & No & No & Yes & No & Yes & Yes \\
\hline
Fig. 5.3 & a & b & c & --- & d & e & f & --- & g \\
\hline
\end{tabular}

\medskip

\begin{tabular}{|c|c|c|c|c|c|c|c|c|c|c|}
\hline
$\lambda$ & 1.01 & 1.1 & 1.225 & 1.25 & 1.3 & 1.5 & 1.6 & 2 & 5 & 10 \\
\hline
Period & HSL & HSL & HSL & 5 & 3 & 1 & 1 & 1 & 1 & 1 \\
\hline
Symmetry & --- & --- & --- & Yes & Yes & Yes & Yes & Yes & Yes & Yes \\
\hline
Fig. 5.3 & --- & h & --- & i & j & --- & k & --- & m & --- \\
\hline
\end{tabular}
\end{center}

For $\lambda \leq 0.533$ we have no
singularities inside $\tilde{C}$, hence the motion is periodic
with the basic period $T=1$, and moreover antiperiodic with period $T=1/2$, so that
all trajectories are symmetrical (see Figs.~5.3a,b). A transition occurs at some
value of $\lambda$ between $0.533$ and $0.8$,
due to a collision among particles $3$ (blue) and $4$ (yellow),
occurring at a time of the order of, or maybe a bit less than, $(3/10)T=3/10$
(see Figs.~5.3b and~5.3c, where clearly $t_{1}=T/10=1/10$). Another transition occurs for a
value of $\lambda$ greater than $0.9687$ but less than $0.98$,
and it causes a branch point to enter
inside $\tilde{C}$ at some value of $\lambda$ between $0.8$ and $0.9687$,
opening thereby the way to a third
sheet of the Riemann surface associated with the solution $\underline{\zeta}(\tau)$
of~(2.5) with~(5.5). This
second transition is due to a~collision, in this case among particles $2$ (green) and $4$
(yellow), occurring at a time of the order, or maybe a little less, than $(8/10)T=8/10$
(see Fig.~5.1d, where clearly again $t_{1}=T/10=1/10$). Then two more transitions occur
(see Fig.~5.3e,f,g)~--- one for~$\lambda$ between $0.98$ and $0.985$,
the other for $\lambda$ between $0.985$
and $0.99$ --- each opening the way to one more sheet of the Riemann surface associated with
the solution $\underline{\zeta}(\tau)$ of~(2.5) with~(5.5) (recall the discussion of Section~4:
overall access
to $P$ sheets including the main one, $P=1+B$ in the notation of Section~4, yields
a \textit{completely periodic symmetrical} motion with period $P$ if $P$
is \textit{odd~--- with every sheet
being visited twice in each period}; it yields a \textit{completely periodic unsymmetrical} motion
with period $P/2$ if $P$ is \textit{even~--- with every sheet being visited once in each period}:
hence for the trajectories shown in Figs.~5.3a,b $P=1$, for those shown in Figs.~5.3c,d $P=2$,
for those shown in Fig.~5.3e $P=3$, for those shown in Fig.~5.3f $P=4$, for those shown in
Fig. 5.3g $P=5$). A more dramatic transition occurs for a value of
$\lambda$ between $1$ and $1.01$,
and it causes a major increase in the complication of the motion, perhaps a transition to
chaos (see Fig. 5.3h). Like in the first example, and of course due to the same mechanism
as discussed above, further increases of $\lambda$ ($\lambda > 1.225$)
cause instead a \textit{decrease} of the
complication of the trajectories, characterized first of all by a return to periodic
motions, and subsequently by a progressive decrease of the period (see Figs.~5.3i--m).

The fourth and last example we report is characterized by the following parameters:
\begin{subequations}
\begin{gather}
N=4 ; \qquad  g_{12}=g_{21}=5, \qquad g_{13}=g_{31}=10, \qquad
 g_{14}=g_{41}=20, \nonumber\\
g_{23}=g_{32}=-20, \qquad g_{24}=g_{42}=-10, \qquad g_{34}=g_{43}=-5,
\label{eq52}
\end{gather}
and by the following values of the parameters $x_{n}^{(0)}$,
$y_{n}^{(0)}$, $\dot{x}_{n}^{(0)}$, $\dot{y}_{n}^{(0)}$
characterizing the initial data via~(5.2):
\begin{gather}
x_{1}^{(0)}=1, \qquad  y_{1}^{(0)}=-2, \qquad \dot{x}_{1}^{(0)}=0,
\qquad \dot{y}_{1}^{(0)}=-1, \nonumber\\
x_{1}^{(0)}=-2, \qquad y_{1}^{(0)}=1,\qquad \dot{x}_{1}^{(0)}=-1,
\qquad \dot{y}_{1}^{(0)}=-1, \nonumber\\
x_{1}^{(0)}=-2, \qquad y_{1}^{(0)}=0 ,\qquad \dot{x}_{1}^{(0)}=0, \qquad
 \dot{y}_{1}^{(0)}=1, \nonumber\\
x_{1}^{(0)}=0, \qquad y_{1}^{(0)}=-2, \qquad \dot{x}_{1}^{(0)}=1, \qquad \dot{y}_{1}^{(0)}=0.
\label{eq53}\end{gather}
\end{subequations}
A representative selection of our findings is reported in Table~5.4, and
some of the corresponding trajectories are displayed in the Figs.~5.4 as indicated there.

\begin{center}
\small

Table 5.4
\medskip

\begin{tabular}{|c|c|c|c|c|c|c|c|c|c|}
\hline
$\lambda$ & 0.5 & 0.736 & 0.8 & 0.894 & 0.9 & 0.904 & 0.905 & 0.91 & 0.95 \\
\hline
Period & 1 & 1 & 1 & 1 & 2 & 2 & HSL & HSL & HSL \\
\hline
Symmetry & Yes & Yes & No & No & No & No & --- & --- & --- \\
\hline
Fig. 5.4 & a & b & c & d & e & --- & --- & --- & --- \\
\hline
\end{tabular}

\medskip

\begin{tabular}{|c|c|c|c|c|c|c|c|c|c|c|}
\hline
$\lambda$ & 1 & 1.1 & 1.125 & 1.13 & 1.15 & 1.2 & 1.5 & 2 & 5 & 10 \\
\hline
Period & HSL & HSL & 2 & 2 & 1 & 1 & 1 & 1 & 1 & 1 \\
\hline
Symmetry & --- & --- & No & No & No & No & No & No & No & No \\
\hline
Fig. 5.4 & f & --- & --- & g & --- & h & --- & --- & i & --- \\
\hline
\end{tabular}
\end{center}

Since the qualitative picture is analogous to those displayed above,
we limit our presentation to the display of Table~5.4 and of the corresponding
set of Figs.~5.4, letting to the alert reader the fun to repeat the
discussion as given above.

\section{Outlook}

The results proven in this paper provide an additional explicit instance of
a phenomenon whose rather general scope has been already advertized via a
number of other examples, treated elsewhere in more or less complete detail
[1--4].

An analogous treatment to that given in this paper will be published soon [5]
in the~con\-text of the ``generalized goldfish model'' [8, 2], which is somewhat
richer inasmuch as it features branch points the nature of which depends on the
values of the coupling constants (in contrast to the case treated herein, where
all relevant branch points are of square-root type), and has moreover the advantage
that its treatment in the $C_{N}$ context is directly interpretable as a genuine (i.e.,
rotation-invariant) real many-body problem \textit{in the plane} (as discussed at the end of
the introductory Section~1, this is not the case for the model considered herein~---
although there does exist a modified version of it in which this ``defect'' is
eliminated~[9]). On the other hand the class of many-body problems with inverse-cube
interparticle potentials, as considered herein, have been (especially, of course, in
the integrable version with equal coupling constants) much studied over the last
quarter century, while the ``goldfish'' model has not yet quite acquired a comparable
``classical'' status.

Of course a more detailed, if perhaps less ``physical'', understanding of
the dynamics of the model studied herein might be gained by investigating numerically
the solutions $\underline{\zeta}(\tau)$ of the equations of motion (2.5) as functions of the
\textit{complex} variable $\tau$ (rather than the solutions $\underline{z}(t)$ as functions of the
\textit{real} variable $t$), and in particular
by mapping out the detailed shape of the multi-sheeted Riemann
surfaces associated with these solutions; and analogous considerations apply to all
the models [1--5] in which the ``trick'', see Section~2, plays a key role. This remains
as a task for the future.

\subsection*{Acknowledgments}

While the results reported in this and related papers were under development we
discussed them with several colleagues, who often provided precious suggestions.
In particular we like to thank for these Robert Conte, Ovidiu Costin, Herman Flaschka,
Giovanni Gallavotti, Nalini Joshi, Martin Kruskal, Fran\c{c}ois Leyvraz, Orlando Ragnisco
and Alexander Turbiner.

\renewcommand{\theequation}{A.\arabic{equation}}
\setcounter{equation}{0}
\section*{Appendix A: the \textit{ansatz} (3.2) }

In this appendix we show that the \textit{ansatz} (3.2) is
consistent with the ODEs (\ref{eq11}), and that the requirement
that (3.2) satisfy (\ref{eq11}) allows to determine uniquely the
coefficient $\beta $ and in principle as well all the coefficients
$\alpha _{l}^{(n)} $ featured by this \textit{ansatz}, while the
remaining coefficients, namely the $4$ constants $b$, $v$, $a$ and
$\tau _{b} $ in (\ref{eq28}), (\ref{eq29}), and the $2(N - 2)$
constants $b_{n}$, $v_{n} $ in (\ref{eq30}), remain
\textit{arbitrary} (except for the obvious requirements $b_{n} \ne
b$, $\,b_{n} \ne b_{m} $, which are hereafter assumed to hold).

Indeed the insertion of (3.2) in (\ref{eq11}) with $n = 1$
respectively $n = 2$ yields
\begin{subequations}
\begin{gather}
 -\frac{1}{4}\beta(\tau - \tau_{b})^{-3/2} +\frac{3}{4}a(\tau -
\tau_{b})^{-1/2} +\sum_{l = 4}^{\infty}\frac{l(l -2)}{4}
\alpha_{l}^{(1)}(\tau-\tau_{b})^{(l - 4)/2}\nonumber\\
\qquad {} =g_{12}(2\beta)^{-3}(\tau-\tau_{b})^{-3/2}
\left\{1+\frac{a}{\beta}(\tau-\tau_{b}) + \sum_{l =
4}^{\infty}\frac{\alpha_{l}^{(1)}-\alpha_{l}^{(2)}}{2\beta}
(\tau-\tau_{b})^{(l-1)/2} \right\}^{-3} \nonumber\\
\qquad {}+\sum_{m = 1,\; m \ne n}^{N} g_{1m}(b-b_{m})^{-3}
\Bigg\{1+(b-b_{m})^{-1} \Bigg[\beta(\tau-\tau_{b})^{1/2}+
(v-v_{m})(\tau-\tau_{b})\nonumber\\
\qquad {} +a(\tau-\tau_{b})^{3/2}  +\sum_{l =
4}^{\infty}\left(\alpha_{l}^{(1)}-\alpha_{l}^{(m)}\right)(\tau-\tau_{b})^{l/2}\Bigg]
\Bigg\}^{-3}\label{eqA1a}
\end{gather}
respectively \begin{gather}
 \frac{1}{4}\beta(\tau - \tau_{b})^{-3/2}
-\frac{3}{4}a(\tau - \tau_{b})^{-1/2} +\sum_{l =
4}^{\infty}\frac{l(l - 2)}{4} \alpha_{l}^{(2)}(\tau-\tau_{b})^{(l - 4)/2} \nonumber\\
\qquad {}=-g_{12}(2\beta)^{-3}(\tau-\tau_{b})^{-3/2} \left\{
1+\frac{a}{\beta}(\tau-\tau_{b}) + \sum_{l =
4}^{\infty}\frac{\alpha_{l}^{(1)}-\alpha_{l}^{(2)}}{2\beta}
(\tau-\tau_{b})^{(l-1)/2} \right\}^{-3} \nonumber\\
\qquad +\sum_{m = 1,\; m \ne n}^{N} g_{2m}(b-b_{m})^{-3} \Bigg\{
1+(b-b_{m})^{-1} \Bigg [-\beta(\tau-\tau_{b})^{1/2}+
(v-v_{m})(\tau-\tau_{b})\nonumber
\end{gather}
\begin{gather}
\qquad {}-a(\tau-\tau_{b})^{3/2} +\sum_{l =
4}^{\infty}\left(\alpha_{l}^{(2)}-\alpha_{l}^{(m)}\right)(\tau-\tau_{b})^{l/2}\Bigg]
\Bigg\}^{-3}. \label{eqA1b}
\end{gather}
\end{subequations}

We now equate in each of these two equations the coefficients of
the term $(\tau - \tau _{b})^{ - 3/2}$ and we get the (same)
relation that determines the coefficient $\beta $,
\begin{equation}
\beta = ( g_{12} /2)^{1/4}.
\end{equation}
And we also see that, with this assignment of $\beta $, in both
equations, (\ref{eqA1a}), (\ref{eqA1b}), the terms that multiply
$(\tau - \tau _{b})^{- 1/2}$ also match exactly.

 It is also easily seen that the remaining
terms can be matched recursively (by expanding the right-hand
sides of these equations, (\ref{eqA1a}), (\ref{eqA1b})~--- as well
as those of (\ref{eqA1c}), see below~--- in powers of $( \tau -
\tau _{b})^{1/2}$), and that in this manner the coefficients
$\alpha _{l}^{(1)}$, $\alpha _{l}^{(2)} $ get uniquely determined,
for instance
\begin{subequations}
\begin{equation}
\alpha _{4}^{(j)} = \sum\limits_{m = 3}^{N} [ ( 13g_{jm} - 3g_{j +
1,m})/20 ]( b - b_{m} )^{ - 3},\qquad j = 1,2\,\,\mbox{mod}(2) .
\end{equation}
\end{subequations}

To complete the analysis one must also check the remaining
equations (\ref{eq11}), with $n > 2$. Insertion of the
\textit{ansatz} (3.2) in these yields \setcounter{equation}{0}
\begin{subequations}
\setcounter{equation}{2}
\begin{gather}
\sum\limits_{l = 4}^{\infty}  \frac{l( l - 2)}{4}\alpha _{l}^{(n)}
(\tau - \tau _{b})^{( l - 4)/2}\nonumber\\
\qquad =g_{n1}(b_{n}-b)^{-3} \Bigg\{ 1+(b_{n}-b)^{-1}
\Bigg[\beta(\tau-\tau_{b})^{1/2}+
(v_{n}-v)(\tau-\tau_{b}) \nonumber\\
\qquad {}+a(\tau-\tau_{b})^{3/2}+\sum\limits_{l = 4}^{\infty}
\left( \alpha_{l}^{(n)} - \alpha_{l}^{(1)} \right)( \tau - \tau
_{b})^{l/2}\Bigg]
\Bigg\}^{- 3} \nonumber\\
\qquad {}+g_{n2}(b_{n}-b)^{-3} \Bigg\{1+(b_{n}-b)^{-1}
\Bigg[\beta(\tau-\tau_{b})^{1/2}+
(v_{n}-v)(\tau-\tau_{b})\nonumber\\
\qquad {}+a(\tau-\tau_{b})^{3/2} +\sum\limits_{l = 4}^{\infty}
\left( \alpha _{l}^{( n)} - \alpha _{l}^{(2)} \right)(\tau - \tau
_{b})^{l/2} \Bigg]\Bigg\}^{ - 3}\nonumber\\
\qquad {}+\sum\limits_{m = 1,\; m \ne n}^{N} g_{nm} (b_{n} -
b_{m})^{- 3}\Bigg\{ 1 + (b_{n} - b_{m})^{- 1}\Bigg[(v_{n} - v_{m})
( \tau - \tau _{b} ) \nonumber\\
\qquad {} +\sum\limits_{l = 4}^{\infty}  \left( \alpha _{l}^{(n)}
- \alpha _{l}^{(m)} \right)(\tau - \tau _{b})^{l/2}
\Bigg]\Bigg\}^{ - 3},\qquad n = 3,4,\ldots,N  . \label{eqA1c}
\end{gather}
\end{subequations}
The consistency of these equations is plain, as well as the
possibility they entail to compute in principle~--- in conjunction
with (\ref{eqA1b}), (\ref{eqA1c})~--- the coefficients $\alpha
_{l}^{(n)} $. For instance one immediately obtains
\setcounter{equation}{2}
\begin{subequations}\setcounter{equation}{1}
\begin{gather}
\alpha_{4}^{(n)}=\frac{1}{2} \left[(g_{n1}+g_{n2})(b_{n}-b)^{-3} +
\sum_{m = 3,\; m \ne n}^{N}
g_{nm}(b_{n}-b_{m})^{-3}\right],\nonumber\\
\qquad \qquad  n = 3,4,\ldots,N .
\end{gather}
\end{subequations}

The treatment of the multiple-collision case, see (\ref{eq27}),
via the corresponding \textit{ansatz}, (3.3), is closely
analogous, but in this case the matching of the coefficients of
the terms proportional to $(\tau - \tau _{b})^{ - 3/2}$
respectively to $(\tau - \tau _{b} )^{- 1/2}$ yields
\begin{subequations}
\begin{gather}
\beta _{n} = - 4\sum\limits_{m = 1,\; m \ne n}^{M}  g_{nm} (\beta
_{n} - \beta _{m})^{ - 3},\qquad n = 1,2,\ldots,M,\\
a_{n} = - 4\sum\limits_{m = 1,\; m \ne n}^{M} g_{nm} (\beta _{n} -
\beta _{m})^{ - 4}(a_{n} - a_{m}),\qquad n = 1,2,\ldots,M.
\end{gather}
\end{subequations}
So in this case the coefficients $\beta _{n}$ and $a_{n}$ get
fixed by these algebraic equations (up, for the coefficients
$a_{n}$, to a common factor). But the basic conclusion about the
\textit{square-root} nature of the branch points is confirmed, as
indeed entailed by the \textit{ansatz} (3.3).

And it can be easily verified that analogous conclusions also
obtain in the case of more general multiple collisions in which
the coordinates collide in groups, for instance \begin{gather}
\zeta _{1} (\tau _{b}) = \zeta _{2} (\tau _{b}),\qquad \zeta _{3}
(\tau _{b}) = \zeta _{4} (\tau _{b}) = \zeta _{5} (\tau
_{b}),\qquad \zeta _{n} (\tau _{b}) \ne \zeta _{1} (\tau _{b}),\nonumber\\
 \zeta _{n}
(\tau _{b}) \ne \zeta _{3} ( \tau _{b}),\qquad \zeta _{n} (\tau
_{b}) \ne \zeta _{m} ( \tau _{b}),\qquad n,m = 6,7,\ldots,N,
\end{gather}
the appropriate \textit{ansatz} being in this case of type
(\ref{eq31}) for the colliding coordinates $\zeta _{n} (\tau)$, $n
= 1,2,\ldots,5$ (with the same values of the constants $b$ and $v$
for each subgroup of colliding coordinates, say, in self-evident
notation, $b_{1} = b_{2}$, $b_{3} = b_{4} = b_{5}$, $v_{1} = v_{2}
$, $v_{3} = v_{4} = v_{5} $), of type (\ref{eq32}) for the non
colliding coordinates $\zeta _{n} (\tau)$, $n = 6,7,\ldots,N$.

\renewcommand{\theequation}{B.\arabic{equation}}
\setcounter{equation}{0}
\section*{Appendix B: the three-body case}

In this appendix we discuss the solution of the evolution
equations (\ref{eq11}) in the \textit{three-body} case, $N = 3$.
As we show below the solution of this three-body problem can be
reduced to quadratures. This fact was discovered by C~Jacobi
almost two centuries ago (but this finding was only published
after his death [10]), was then rediscovered by C~Marchioro (who
also did not publish this finding) and was finally neatly
presented by D~C~Khandekar and S~V~Lawande [11], who actually
treated (\ref{eq4}) rather than (\ref{eq11}) (they were not aware
at the time of the ``trick'' (2.4) that relates (\ref{eq11}) to
(\ref{eq4}), and they mainly focussed on the integrable case with
equal coupling constants and on its connection with the
corresponding quantal problem~--- for a bit more on the history of
this problem see Section~2.N of Ref.~[2]).

So we now focus on the evolution equations
\begin{equation}
\label{eqB1} \zeta''_{n} = \sum\limits_{m = 1, \; m \ne n}^{3}
g_{nm} ( \zeta _{n} - \zeta _{m})^{- 3},\qquad n = 1,2,3,\quad
g_{nm} = g_{mn} ,\quad\zeta _{n} \equiv \zeta _{n} (\tau),
\end{equation}
which clearly obtain from the Hamiltonian
\begin{subequations}
\begin{gather}
H = \frac 12 \left( p_{1}^{2} + p_{2}^{2} + p_{3}^{2} \right)
+\frac 12 \left[ g_{12} (\zeta_{1}-\zeta_{2})^{-2} + g_{23}
(\zeta_{2}-\zeta_{3})^{-2} + g_{31} (\zeta_{3}-\zeta_{1})^{-2}
\right]\label{eqB2a}
\end{gather}
hence entail that the quantity \begin{gather}
 H = \frac 12 \left(
\zeta^{\prime2}_{1} + \zeta^{\prime 2}_{2} + \zeta^{\prime 2}_{3}
\right)
 + \frac 12 \left[ g_{12} (\zeta_{1}-\zeta_{2})^{-2}\! +
g_{23} (\zeta_{2}-\zeta_{3})^{-2}\! + g_{31}
(\zeta_{3}-\zeta_{1})^{-2} \right]\!\label{eqB2b}
\end{gather}
\end{subequations}
is a constant of the motion. It is also plain that the initial
position, $\bar {\zeta} (0)$, and the (initial) velocity, $V$, of
the center of mass
\begin{equation}
\bar{\zeta}  = (\zeta _{1} + \zeta _{2} + \zeta _{3})/3,
\end{equation}
are two additional constants of motion, since the evolution
equations (\ref{eqB1}) clearly entail that the center of mass
moves uniformly:
\begin{equation}
\label{eqB4} \bar{\zeta} (\tau) = \bar{\zeta} (0) + V\tau.
\end{equation}
And we moreover now know (see the last part of Section 2) that the
quantity \begin{subequations}
\begin{equation}
\label{eqB5a} \bar {\zeta} ^{(2)} = \left( \zeta _{1}^{2} + \zeta
_{2}^{2} + \zeta _{3}^{2}  \right)/3
\end{equation}
evolves according to the simple equation (see (\ref{eq18}))
\begin{equation}
\label{eqB5b} (\bar{\zeta}^{(2)})''=(4/3)H,
\end{equation}
\end{subequations}
with $H$ defined by (\ref{eqB2b}).

It is now convenient [11] to introduce the ``Jacobi coordinates''
\begin{subequations}
\begin{equation}
\label{eqB6a} \eta = 2^{ - 1/2}(\zeta _{1} - \zeta _{2}),\qquad
\xi = 6^{ - 1/2}(\zeta _{1} + \zeta _{2} - 2\zeta _{3}),
\end{equation}
and moreover to introduce the corresponding ``circular
coordinates'' by setting
\begin{equation}
\label{eqB6b} \eta = \rho \sin( \theta),\qquad \xi = \rho \cos(
\theta).
\end{equation}
\end{subequations}
Note that these definitions entail
\begin{subequations}
\begin{gather}
\label{eqB7a}
\zeta _{1} = \bar {\zeta}  + 2^{ - 1/2}\eta + 6^{ - 1/2}\xi = \bar
{\zeta}  -( 2/3)^{1/2}\rho \cos[ \theta + (2\pi /3)] ,\\
\label{eqB7b}
\zeta _{2} = \bar {\zeta}  - 2^{- 1/2}\eta + 6^{ - 1/2}\xi = \bar
{\zeta}  - (2/3)^{1/2}\rho \cos[\theta -(2\pi /3)],\\
\label{eqB7c}
\zeta _{3} = \bar {\zeta}  - (2/3)^{1/2}\xi = \bar {\zeta
} - (2/3)^{1/2}\rho \cos\theta,
\end{gather}
\end{subequations}
as well as
\begin{subequations}
\begin{gather}
\label{eqB8a}
H = (3/2)V^{2} + (1/2){\rho} '^{2} +
(1/2)\rho ^{2}{\theta} '^{2} + ( 1/4) \rho ^{ - 2}f(\theta),\\
\label{eqB8b}
f(\theta) = g_{12} /\sin^{2}(\theta) +
g_{23}/\sin^{2} \left[ \theta + (2 \pi/3) \right] +
g_{31} /\sin^{2} \left[ \theta - (2 \pi/3) \right],
\end{gather}
\end{subequations}
and
\begin{equation}
\bar {\zeta} ^{(2)} = \bar {\zeta} ^{2} + (1/3)\rho ^{2} .
\end{equation}

From the last formula and (\ref{eqB4}), (\ref{eqB5b}) we get
\begin{subequations}
\begin{equation}
\label{eqB10a}
\left( \rho^{2} \right)'' = 4H - 6V^{2} ,
\end{equation}
hence
\begin{gather}
\label{eqB10b}
\rho ^{2}(\tau) = A\left( \tau ^{2} + 2B\tau + C \right) =
A(\tau - \tau _{ +})(\tau - \tau _{ -}),\\
\label{eqB10c}
A = 2H - 3V^{2},\\
\label{eqB10d}
\tau _{ \pm}  = - B \pm D,\\
\label{eqB10e}
D^{2} = B^{2} - C.
\end{gather}
\end{subequations}
Here we have introduced two other constants of integration, $B$ and $C$,
while the constants~$A$ and $D$ are defined by (\ref{eqB10c}) and (\ref{eqB10e}).

We now introduce this expression of $\rho(\tau)$ in (\ref{eqB8a})
and easily obtain the formula
\begin{subequations}
\begin{equation}
\label{eqB11a}
{\theta} '^{2} = - \left[ A^{2} D^{2} + (1/2) f (\theta) \right] \rho ^{ - 4},
\end{equation}
that entails the quadrature
\begin{equation}
\label{eqB11b}
\int^{\theta}   d{\theta} ' \left[  -A^{2}D^{2} -
(1/2) f({\theta} ') \right]^{ -1/2} = \int^{\tau}   d{\tau} ' \left[ \rho ( \tau ') \right]^{ - 2}.
\end{equation}
\end{subequations}

The integral in the right-hand side of this formula, (\ref{eqB11b}), is easily done
(see (B.10)), and one arrives thus at the final formula
\begin{subequations}
\begin{gather}
\label{eqB12a}
F(\theta) = \mbox{arccotan}\, [(\tau + B)/D],\\
\label{eqB12b}
F(\theta) = \int^{\theta}  d{\theta} '\left[
1 + (1/2) (AD)^{-2} f(\theta ') \right]^{ - 1/2} ,
\end{gather}
\end{subequations}
where of course $f(\theta)$ is defined by (\ref{eqB8b}).

In the \textit{integrable} ``equal-coupling-constants'' case,
\begin{subequations}
\begin{equation}
\label{eqB13a}
g_{nm} = g,
\end{equation}
it is easily seen that
\begin{equation}
\label{eqB13b}
f(\theta) = 9g[\sin(3\theta)]^{ - 2},
\end{equation}
hence the integration in the right-hand side of (\ref{eqB12b}) is easily performed
to yield
\begin{equation}
\label{eqB13c}
F(\theta) = (1/3)\{ E -\arcsin[ \cos(3\theta)/\mu]\},
\end{equation}
where $E$ is an integration constant and
\begin{equation}
\label{eqB13d}
\mu = \left[ 1 + (9/2)g(AD)^{-2}\right]^{1/2}.
\end{equation}
\end{subequations}
Hence in this case one gets for $\theta(\tau)$ the
completely explicit expression
\begin{equation}
\theta =(1/3)\arccos[\mu \sin\{
E-3\,\mbox{arccotan}\,[ (\tau + B)/D]\}],
\end{equation}
which, together with (\ref{eqB10b}) and (B.4), provides via (B.7) completely
explicit expressions of the coordinates $\zeta _{n} (\tau)$,
$n = 1,2,3$. From these it is immediately seen that the
coordina\-tes~$\zeta _{n} (\tau)$, considered as functions of
the complex variable $\tau $, feature two kinds of \textit{square-root}
branch points: those occurring at $\tau = \tau _{ \pm}  $, see (\ref{eqB10b}), (\ref{eqB10d}),
which clearly correspond to \textit{triple} collisions, see (\ref{eqB10b}) and (B.7);
and those occurring at $\tau = \tau _{b} $,
\begin{equation}
\label{eqB15}
\tau _{b} = - B + D\mbox{cotan}\{(1/3)[E-\arcsin(1/\mu)+(2\pi/3)k]\},
\qquad k = 0,1,2,
\end{equation}
which correspond instead to \textit{pair} collisions. Indeed it is easily
seen that, for $\tau \approx \tau _{b} $,
\begin{subequations}
\begin{gather}
\label{eqB16a}
\theta = (2\pi /3)k + \alpha [(\tau - \tau_{b})/D]^{1/2} + O(|\tau - \tau _{b}|),\qquad k = 0,1,2,\\
\label{eqB16b}
\alpha=i(2/3)^{1/2}\left[ 1+(2/9)(AD)^{2}/g \right]^{-1/4}
D\left[ (\tau_{b}+B)^{2}+D^{2} \right]^{-1/2};
\end{gather}
\end{subequations}
and (\ref{eqB16a}) clearly entails, see (B.7), that a \textit{pair} collisions
occurs at $\tau = \tau _{b} $.

The analysis of the periodicity of the solutions of (2.1) in this case is
plain; of course the fact that the number of branch points is finite entails
that in this \textit{integrable} case \textit{all nonsingular} solutions are
\textit{completely periodic}, either with period~$T$, see (\ref{eq6}), or with a
period which is a (small) entire multiple of~$T$, confirming the results
implied by the general treatment~[2] of the \textit{integrable}
equal-coupling-constant $N$-body problem~(2.1).

In the general case with \textit{arbitrary} coupling constants one can
rewrite (\ref{eqB12b}) as follows (by setting $u = 4\sin^{2}{\theta} ' - 3$):
\begin{subequations}
\begin{gather}
\label{eqB17a}
F(\theta)=(1/2)\int^{U(\theta)}
du \, u(1-u)^{-1/2}\nonumber\\
\phantom{F(\theta)=}{}\times \left[u^{2}(u+3+\tilde{G})+G
+\hat{G}(3+u)^{3/2}(1-u)^{1/2}\right]^{-1/2},\\
\tilde{G}=(2g_{12}-g_{23}-g_{31})(AD)^{-2} ,\nonumber\\
\label{eqB17b}
G=9(g_{23}+g_{31})(AD)^{-2}, \qquad\hat G= 3^{1/2}(g_{23}-g_{31})(AD)^{-2},\\
\label{eqB17c}
U(\theta) = 4[\sin\theta]^{2} -3 .
\end{gather}
\end{subequations}
In the equal-coupling-constants case, see (\ref{eqB13a}), things simplify because
\begin{subequations}
\begin{equation}
\tilde {G} = \hat {G} = 0,
\end{equation}
see (\ref{eqB17b}), hence the integral in the right-hand side of (\ref{eqB17a}) can then
be performed in terms of elementary functions, since via the change of
variable
\begin{equation}
u^{2}(u + 3) = 4v,
\end{equation}
somewhat miraculously (albeit not surprisingly) (B.17) yield
\begin{gather}
F(\theta) = (1/3)\int^{W(\theta)} dv\,\left\{ ( 1 - v)(G + v) \right\}^{- 1/2},
\\
W(\theta) = [\sin(3\theta)]^{2};
\end{gather}
\end{subequations}
and of course one gets thereby again (\ref{eqB13c}).

If instead (only) the two coupling constants $g_{23} $ and $g_{31} $
coincide, $g_{23} = g_{31} $, entailing $\hat {G} = 0$, the integral in the
right-hand side of (\ref{eqB17a}) is of elliptic type (of course the same outcome,
namely the elliptic character of the solutions~[9], characterizes any case
with two equal coupling constants, although this is not immediately evident
from~(\ref{eqB17a})). But not much enlightenment seems to obtain, even in this
(somewhat simpler) case, from an attempt to perform explicitly the
integration in the right-hand side of (\ref{eqB17a}).

One must rather try and evince information directly from the integral
representation of $F(\theta)$, see (\ref{eqB12b}) with (\ref{eqB8b}), or (B.17),
or from the following representation which the diligent reader will have no difficulty
in deriving:
\begin{subequations}
\begin{gather}
F(\theta)=(2i)^{-1}  \int^{\exp(2 i \theta)}  dw \, w^{-1} \left(w^{3}-1\right)
[ \phi(w)]^{-1/2} ,\\
\phi(w)=\left(w^{3}-1\right)^{2}-4(AD)^{-2} w \Big\{
g_{12} [ w-\exp(2 \pi i/3)]^{2}
[ w-\exp(-2 \pi i/3)]^{2}\nonumber\\
\phantom{\phi(w)=} {}+(w-1)^{2} \left[
g_{23}[ w-\exp(-2 \pi i/3)]^{2}+
g_{31} [ w-\exp(2 \pi i/3) ]^{2} \right] \Big\} .
\end{gather}
\end{subequations}
It is now clear from these formulas, together with (B.7), (B.10)
and (\ref{eqB12b}), that~--- just as in the \textit{integrable}
equal-coupling-constants case~--- also in the case with
\textit{arbitrary} coupling constants the solutions $\zeta _{n}
(\tau)$ feature two kinds of \textit{square-root} branch points:
those associated with \textit{triple} collisions, which occur at
$\tau = \tau _{ \pm}  $, see (\ref{eqB10b}), (\ref{eqB10d}), and
those associated with the vanishing of the derivative
$F'(\theta)\equiv dF(\theta)/d\theta$ of $F(\theta)$, namely those
associated with the $3$ values $\theta = 0$, $\theta = 2\pi /3$,
$\theta = - 2\pi /3$ (see (B.19); of course $\theta $ is defined
$\mbox{mod}(2\pi)$, see (\ref{eqB6b})), which of course correspond,
as expected, to \textit{pair} collisions, see (B.7). It is however
not possible now~--- in contrast to the \textit{integrable} case,
see (\ref{eqB15})~--- to obtain an explicit expression for the
values $\tau _{b} $ of $\tau $ at which these branch points occur:
but we infer from (B.17) and (\ref{eqB12a}) that generally there
is an \textit{infinite} number of such branch points in the
complex $\tau $-plane (again, in contrast to the
\textit{integrable} case, when there only are \textit{three}, see
(\ref{eqB15})). It is on the other hand easy to obtain from these
formulas an explicit expression for the behavior of $\theta
(\tau)$ for $\tau \approx \tau _{b} $, for instance for the branch
point with $\theta (\tau _{b}) = 0$,
\begin{subequations}
\begin{gather}
\theta(\tau)=\beta [(\tau-\tau_{b})/A]^{1/2}+O(|\tau-\tau_{b}|),\\
\beta=i(b_{12}/2)^{1/4} \left[ D^{2}+(\tau_{b}+B) \right]^{-1/2}.
\end{gather}
\end{subequations}

\label{calogero-lastpage}

\end{document}